\documentclass{aa}

\usepackage{txfonts}
\usepackage{graphicx}
\usepackage{amsmath}
\usepackage{color}
\usepackage{gensymb}

\bibpunct{(}{)}{;}{a}{}{,} 

\usepackage{hyperref}

\newcommand{\equ}[1]{Eq.~(\ref{#1})}
\newcommand{\fig}[1]{Fig.~\ref{#1}}
\newcommand{\Fig}[1]{Figure~\ref{#1}}
\newcommand{\figs}[2]{Figs.~\ref{#1}-\ref{#2}}
\newcommand{\Figs}[2]{Figures~\ref{#1}-\ref{#2}}

\newcommand{\f}[2]{\frac{#1}{#2}}
\newcommand{\rsun}{R_\odot}

\begin{document}
\title{Observational characterisation of large-scale 
transport and horizontal turbulent diffusivity in the quiet Sun}
\titlerunning{Observational characterisation of horizontal turbulent transport in the quiet Sun}
\author{F. Rincon\inst{1,2}\and  P. Barr\`ere\inst{1,2,3} \and
  T. Roudier\inst{1,2}}
\institute{CNRS; IRAP; 14 avenue Edouard Belin, F-31400 Toulouse,
           France,
\and 
Universit\'e de Toulouse; UPS-OMP; IRAP: Toulouse, France,
\and
Universit\'e Paris-Saclay, Universit\'e Paris Cit\'e, 
           CNRS, AIM, 91191, Gif-sur-Yvette, France\\
\email{frincon@irap.omp.eu}}

\date{\today}

\abstract{The Sun is a magnetic star, and the only 
  spatio-temporally resolved astrophysical system displaying turbulent magnetohydrodynamic 
  thermal convection. This makes it a privileged object of study to understand 
  fluid turbulence in extreme regimes and its interactions with magnetic fields.
  Global analyses of high-resolution solar observations provided by the NASA Solar 
  Dynamics Observatory (SDO) can shed light on the physical processes 
  underlying large-scale emergent phenomena such as the solar dynamo cycle.
  Combining a Coherent Structure Tracking reconstruction of photospheric flows, 
  based on photometric data, and a statistical 
  analysis of virtual passive tracers trajectories advected by these flows,
  we characterise one of the most important such processes, turbulent diffusion,
  over an unprecedentedly long monitoring period of six consecutive days of
  a significant fraction of the solar disc.
  We first confirm, and provide a new global view of the 
  emergence of a remarkable dynamical pattern of Lagrangian Coherent Structures tiling 
  the entire surface. These structures act as transport barriers on the time and spatial scale of 
  supergranulation and, by transiently accumulating particles and magnetic fields, 
  appear to regulate large-scale turbulent surface diffusion.
  We then further statistically characterise the turbulent transport regime using two different methods,
  and obtain an effective horizontal turbulent diffusivity 
  $D=2-3\times10^8~\mathrm{m}^2~\mathrm{s}^{-1}$ on the longest timescales considered.
  This estimate is consistent with the transport coefficients required in 
  large-scale mean-field solar dynamo models, and is in broad agreement with the results 
  of global simulations. Beyond the solar 
  dynamo, our analysis may have implications for understanding the structural 
  connections between solar-surface, coronal and solar-wind dynamics, and it also provides
  valuable lessons to characterise turbulent transport in other, unresolved 
  turbulent astrophysical systems.}

\keywords{Sun: photosphere -- Convection -- Turbulence -- Magnetic fields -- Dynamo}
\maketitle

\section{Introduction}
\subsection{Context and motivation}
Thermal convection is one of the most common fluid
transport processes encountered in astrophysics, and the Sun and its
photosphere provide us with a unique observationally well-resolved
example of such (magnetohydrodynamic, MHD) turbulence in highly nonlinear
regimes. Indeed, dynamical MHD phenomena on the Sun are now
continuously monitored with temporal and
spatial resolutions of the order of seconds and hundred 
kilometers respectively, which are truly astonishing small numbers by
astronomical standards. As such, the Sun is a special place
to study nonlinear thermal and MHD transport
processes, and to understand how they can affect the structure and
evolution of many astrophysical systems that can not be
resolved by observations, or emulated in laboratory experiments.

While some aspects of solar (MHD) convection, such as solar
granulation, are well understood \citep{nordlund09}, we still
lack definitive answers to many important questions
such as how thermal turbulence in the Sun organises on large
scales, how it interacts with and amplifies magnetic fields at both
large and small-scales, how it transports quantities 
such as angular momentum or magnetic flux
\citep{miesch05,hathaway12,charbonneau14,brun17,rincon18}.
One limitation is that despite much progress in helioseismology, 
we do not (yet) have time and space-resolved determinations of 
internal multiscale convective dynamics in the solar convection zone
\citep{gizon10,hanasoge12,svanda12,duvall13,duvall14,degrave15,
hanasoge15,greer15,toomre15}. 
Photospheric observations still allow for the most direct
characterisation of solar convection and are therefore most helpful
to put observational constraints on dynamical transport
processes, such as turbulent diffusion, which are key to understand
emergent dynamical phenomena such as the large-scale dynamo cycle.
Given the interfacial nature of the photosphere,
understanding magnetic transport there is also critical to understand 
the energetics and magnetic dynamics of the corona \citep[e.g.][]{amari15}, 
and the near-Sun structure of 
the solar wind recently uncovered by the Parker Solar Probe 
\citep{bale19,kasper19}. Finally, characterising
large-scale solar dynamical processes in detail
could provide useful insights into many unresolved similar 
astrophysical (MHD) turbulent processes (e.g. accretion, star formation, 
cosmic ray diffusion, galactic and extragalactic magnetogenesis)
that can not be modelled in full detail due to their
extreme nonlinearity and multiscale essence.

\subsection{Solar-surface velocity measurements and their use}
The most direct observational inference of
multiscale photospheric flows is via measurements of Doppler-projected
velocities \citep{leighton62,rincon18}. Even a simple
visual inspection of Doppler images clearly reveals the pattern of
supergranulation flow ``cells'' \citep{hart54}. 
The trademark signature of the
supergranulation flow is a power excess around $\ell\sim 120-130$
(35~Megameters (Mm)) in the spherical harmonics power spectrum of
global maps of Doppler-projected velocities obtained either with the
MDI instrument aboard SOHO \citep{hathaway2000} or with the HMI
instrument aboard SDO \citep{williams14, hathaway15}.
Dopplergrams can be complemented by other observational inference
techniques, such as Local Correlation Tracking
\citep[LCT,][]{november88} or Coherent Structure Tracking
\citep[CST,][]{rieutord07}, to reconstruct the horizontal components of 
the velocity field.
Both techniques require photometric data with high spatio-temporal
resolution to resolve small-scale structures and motions and are
a bit demanding computationally. In particular, the CST, which 
we use in this paper, requires tracking the motions of
a large statistical ensemble of small-scale intensity structures
(granules) advected by larger-scale flows \citep{rieutord01}.
For this reason, they have for a long time mostly
been applied to limited field-of-views, obtained from either ground-based
\citep{november88,november89,roudier99,rieutord01,rieutord08}, or
space-based observatories such as TRACE and Hinode
\citep{simon04,roudier09, rieutord10b}, although local correlation
tracking has also been applied to larger patches of MDI data
(\cite{shine00,svanda07}, see Fig.~22 of \cite{nordlund09}). 
However, the availability of highly-resolved full-disc photometric 
and Doppler data, as now routinely delivered by SDO/HMI, 
and improved numerical data processing capacities, have opened 
the 24~h/24~h possibility to infer multiscale flows at the photospheric 
level from local to fully global scales.

A combined CST/Doppler analysis of full-disc photometric SDO/HMI
data allowing reconstruction of the three components of the 
velocity field was first attempted by
\cite{roudier12,roudier13}. LCT was also applied by
\cite{langfellner15} to $180\times 180$ Mm$^2$ patches of SDO/HMI
images in order to characterise averaged properties of supergranules. 
A marked improvement on these techniques
was introduced by \cite{rincon17}, hereafter R17,
which made it possible to produce accurate full-disc maps (up to
$60\degree$ from the disc centre) of the Eulerian horizontal and
radial flow components at spatial scales larger than 2.5~Mm, 
with a time cadence of 0.5~h. With this data, R17 
could notably calculate the full spherical harmonics kinetic energy
spectrum of the radial, poloidal, and toroidal components of the
photospheric velocity field over a wide range of horizontal
scales. These global measurements separating different
flow components notably inspired a new theoretical
anisotropic turbulence phenomenology of large-scale photospheric
convection. They can also serve as a consistency check for numerical
simulations, and were shown by \cite{rincon18} to compare well with
global simulations of solar convection \citep{hotta14}. An alternative 
technique for the determination of the full surface velocity field 
over the solar disc based on Doppler data only was recently 
introduced by \cite{kashyap21}.

\subsection{Objectives and approach taken in this study}
The aim of this work is to further exploit the possibilities offered
by the application of the CST to the full-disc, continuous SDO/HMI data stream,
to characterise the large-scale transport and turbulent diffusion properties
of photospheric flows. Because the technique does not give us access
to the vertical dependence of flows, the analysis is necessarily
limited to horizontal transport, which is nevertheless
dominant at the scales considered as a result of the strong flow
anisotropy (R17). In R17, a 24~h sequence of full-disc SDO/HMI data
was used, which was sufficient to calculate simple
statistical quantities such as kinetic energy spectra. In this paper though,
we aim at probing large-scale transport on timescales longer than the 
24-48~h correlation time of supergranules, the most energetic dynamical surface 
structures, and therefore use a significantly longer,
uninterrupted six-day sequence. Characterising the dynamics on 
such a long timescale creates new challenges, 
such as following regions of interest over a substantial fraction 
of the solar rotation period, and it is also limited by the lack of precision 
of the velocity-field deprojection near the solar limb. For this reason, 
a week is about the maximum continuous integration time achievable with our
procedure, but it nevertheless provides us with an unprecedented,
statistically-rich turbulent Eulerian velocity-field dataset.

To characterise turbulent transport, we simulate and statistically analyse the
Lagrangian dynamics of passive particles virtually distributed over the surface
and advected by the inferred, time-evolving Eulerian horizontal
velocity field. This, first of all, enables us to derive global maps of Finite
Time Lyapunov Exponents
(FTLEs), introduced in a solar-surface physics context a few years ago
by \cite{yeates12}, see also \cite{chian14}. Besides some quantitative 
insights into the intrinsic chaoticity of solar surface flows, such an 
analysis enables us to globally map for the first time a network of 
so-called Lagrangian Coherent Structures (LCS), which are the 
loci of transient accumulation or rarefaction of tracers.
Using passive tracer statistics, we then further probe the
transport regime of the flow up to timescales of a week,
and derive an associated large-scale turbulent diffusion coefficient 
at the photospheric level. 

This study bears similarities, and 
shares some  diagnostics with a more common approach to transport 
in solar physics, based on the direct tracking of magnetic
elements in magnetograms. However, due to the ephemeral 
nature, and occasional cancellation of magnetic elements, the latter approach
has traditionally been limited to short-time (a few tens of hours at best)
intra-supergranule transport, i.e. network formation \citep[][see \cite{bellot19} 
for a recent review]{schrijver96,berger98,hagenaar99,utz10,
manso11,abramenko11,orozco12,giannattasio13,giannattasio14,jafarzadeh14,yang15}.
A notable exception is \cite{iida16}, who extended this type of analysis 
to five consecutive days, albeit with very noisy statistics on the longest 
times probed.  While more indirect, our approach offers a way to 
bypass such limitations, as virtual passive tracers can be introduced 
numerically with arbitrary resolution, and followed with more precision, 
ease, better statistics and for significantly longer times 
than magnetic elements.
Based on a comparison of tracer concentrations and trajectories
with HMI magnetograms, we will argue that the transport 
properties thus obtained are representative of those of the 
weak ($< 100$~G), essentially passive, magnetic fields 
at the surface of the quiet Sun. 

Section~\ref{data} introduces the observation and velocity
datasets, and Sect.~\ref{fieldanalysis} the tools of fluid flow
analysis applied to the data. Section~\ref{LCSanalysis} presents
a Lagrangian flow analysis using FTLEs and LCS, and
a comparison between the latter and the magnetic network. 
Sect.~\ref{transport} provides a phenomenological and 
quantitative statistical characterisation of 
turbulent transport and estimates horizontal turbulent coefficient 
using two different methods.  The implications of our work
for the broader understanding of large-scale dynamics 
and turbulent transport are discussed in Sect.~\ref{discussion}.

\section{Data processing\label{data}}

The data used for this study has already been presented in \cite{roudier23} (hereafter R23), 
and the reduction procedures have been described in R17 and R23. We restate the main 
information here for the sake of completeness, outlining the differences with R23 where 
necessary.

\subsection{The data}
Our analysis is based on six days of uninterrupted high-resolution white-light 
intensity and Doppler observations of the entire solar disc by the HMI 
instrument aboard the SDO satellite \citep{scherrer12,schou12}.
The data was obtained from 26 November (00:00:00 UTC) to 1 December 2018 (23:59:15 UTC),

\subsection{Image corrections and derotation}
Different procedures, detailed in App. A of R17, were first applied to the images
to correct for misalignment, change in size of the solar disc, and the limbshift 
effect. In a second step, we adjusted the differential 
rotation profile from the raw Doppler data averaged over one day of observation, 
and used the resulting rotational velocity signal to derotate all images
so as to work in a reference frame corotating with the Sun. 
This way, any given image pixel after derotation corresponds 
to a fixed physical location on the solar surface. 
For the derotation procedure, we used the rotation profile derived by R23,
\begin{equation}
     \Omega(\lambda) = A + B \sin^2\lambda + C \sin^4\lambda,
 \end{equation}
where $\lambda=$ here denotes the latitude, 
$A=2.864 \times 10^{-6}\,\mathrm{rad~s^{-1}}$, $B=-5.214 \times 10^{-7}\,\mathrm{rad~s^{-1}}$, 
and $C=-2.891 \times 10^{-7}\,\mathrm{rad~s^{-1}}$. This corresponds to a velocity 
of 1.9934~km~s$^{-1}$ (14.1781\textdegree per day) at the equator. 

Derotation was applied to the white light intensity data, correcting the mean 
differential rotation of the Sun to bring back longitudes related to 
the solar surface at the same locations for each time deviation 
from the origin of the first HMI image. The reference time for the CST code 
was taken at 00:00  UTC, 29 November 2018, the middle of our sequence  
(see code manual\footnote[1]{\url{https://idoc.ias.u-psud.fr/system/files/user_guide_annex_version1.2_26mars2021.pdf}}).
Accordingly, derotation was applied on 26 to 28 November 
from the left to the right (east to west) and on 29 November to 1 december 
from the right to the left (west to east).

\subsection{Derivation of the Eulerian photospheric velocity field projected in the CCD plane}
We subsequently used the Coherent Structure Tracking (CST) algorithm to derive the projection
($u_x,u_y$) in the CCD plane of the corotating photospheric Eulerian velocity field on scales larger than 
2.5~Mm, tracking ensembles of granules advected by horizontal flows.  These are the only 
components of the flow needed for the Lagrangian tracer analyses performed in this paper (for a 
reconstruction of the full $(u_r,u_\theta,u_\varphi)$ velocity field in spherical coordinates, see R17).

By means of the derotation procedure, the corotating solar disc areas 
under consideration effectively remain centered, throughout the analysis, on an effective 
corotating reference disc centre (chosen as the actual
disc centre in the middle time of the full sequence). This enables
us to compute Lagrangian tracer trajectories 
over the chosen area using the in-CCD-plane 
$(u_x,u_y)$ velocity field. We explain how this is done in Appendix.

\subsection{Masking and apodising\label{datalimitation}}
The determination of velocity fields close to the limb with the CST is 
more noisy and of lesser quality due to projection and resolution effects.
Therefore, we finally apodised the velocity maps obtained with the procedures 
described above to focus on a limited corotating zone of the surface sufficiently
far away from the limb, so that reliable velocity field data could be exploited 
continuously for several days.

In the following, we present results using circular apodising windows of different angular openings 
with respect to the disc centre. The rule of thumb here is that the larger the window, the shorter
the time over which our Lagrangian tracer analyses can be conducted, as larger regions spend less time 
in full visibility on the side of the Sun visible to SDO, and sufficiently far away from the limb. We conducted 
some test analyses to determine which apodising window worked best to track flows reliably over the 
largest timescale possible, and converged to a circular apodising window of $23.5\degree$ 
opening with respect to the disc centre for the longest six-day Lagrangian tracer 
analysis presented in Sects.~\ref{LCSanalysis}-\ref{transport}.

Considering the limitations of the CST to infer flows close to the limb, this masking strategy is the 
only reasonable course of action to ensure that all successive corotating velocity field data snapshots 
used to calculate Lagrangian trajectories are equally reliable. Of course, this limits the available 
corotating fraction of the solar disc used, compared to the full-disc raw intensity observations of SDO, 
and also explains why our analyses are limited to six days at most, rather than twelve-thirteen days 
if we were able to infer flows in a given corotating region reliably from the time that region becomes 
visible on one side of the limb to the time it disappears on the other side.

\subsection{The final data product}
The final data product, used in the following, is a sequence of 288 velocity-field maps 
of a single corotating physical area of the solar surface, tracked over six days, with a 
temporal resolution of 30 min and spatial resolution of 2.5~Mm. The angular opening
of the \textit{minimum} $23.5\degree$ apodising window used in the paper corresponds 
to an arc of 574~Mm at the surface, but larger fields of views were also 
exploited over shorter periods (see Sect.~\ref{LCSanalysis}).  
This minimal $23.5\degree$ apodised field of view still
contains the equivalent of $\sim 350$ supergranules of 30~Mm diameter, that is 50 times 
more than the Hinode field of view. Besides, the continuous monitoring by SDO from space 
enables continuous tracking and transport of structures and tracers on much longer timescales 
(up to six days) than possible with 
ground-based instruments with comparably wide field of views, such as the CALAS camera used at Pic du Midi 
for a similar purpose a few years ago, which at best only provided continuous monitoring over 7.5~h
of a $290\times 216~\mathrm{Mm}^2$ field of view containing $\sim 70$ supergranules \citep{rieutord08}.

Overall, the extents of our field of views and continous monitoring periods ensure 
that a robust, and unprecedented statistical analysis of turbulent transport on temporal 
and spatial scales much larger than those of individual supergranules can be carried 
out with the data.  

\section{Flow analysis techniques\label{fieldanalysis}}
\subsection{Lagrangian analysis of solar surface flows}
\subsubsection{Finite-time Lyapunov exponents (FTLEs)}
The calculation of FTLEs provides a
convenient mathematical way of characterising the chaoticity of
dynamical systems, and in fluid dynamics to investigate the dynamical
Lagrangian properties and organization of a flow. In particular, a
flow exhibiting Lagrangian
chaos (i.e. exponential divergence of the trajectories of initially
close fluid particles) is characterised by the fact that at least one
of its Lyapunov exponents is positive. To be more precise, for complex
flows, the value of Lyapunov exponents usually depends on the spatial
location and time,  so instead of a single number, we compute a scalar
field of Lyapunov exponents, which itself evolves dynamically in time. 
Computing a field of FTLEs for a given
fluid flow requires to integrate the trajectories of passive Lagrangian
tracers in the flow for a target time $T$. The principles of this
kind of calculation are extensively documented in the literature
\citep[see e.g.][]{haller01a,shadden05,green07,lekien07,lekien10} and 
are summarised in Appendix~\ref{FTLEapp}. Our own implementation is
very much based on that described by \cite{lekien10}. 

The main specificities of the problem at hand are that it has a 
global spherical, non-Euclidean geometry, and that we only have access
to the flow on a single spherical surface. The latter implies that we
can not compute the three FTLE scalar fields of the full 3D photospheric
flow. However it is still possible to compute two FTLE scalar
fields associated with the horizontal flow on the surface\footnote{%
  The two FTLEs of the flow on the surface are probably close to 
  two of the three FTLEs of the full 3D flow, because the photospheric
  velocity field at the large scales considered in this paper is
  strongly anisotropic with respect to the radial direction (R17).}.
Indeed, as shown by \cite{lekien10}, FTLEs for a 2D flow on a smooth
non-Euclidean surface such as a sphere can be indirectly inferred from
the projection in a 2D plane (in our case, the plane of the sky/CCD) of the
Lagrangian tracers trajectories using a mapping (described in
App.~\ref{mappingFTLEapp}). This is fortunate, because the projected
trajectories are easier to compute than the actual trajectories on a
spherical surface, provided that the projection of the velocity
field in the projection plane is known. This is the case in our
problem: the projection of the photospheric velocity field in the $(x,y)$
plane of the satellite CCD is precisely what the CST algorithm computes.

Our implementation of the FTLE computation algorithm in a 2D
plane, including the integration of Lagrangian tracers trajectories 
described in App.~\ref{tracersFTLEapp}, and subsequent FTLE calculation
described in App.~\ref{2DFTLEapp}, was validated using the chaotic
double-gyre analytical flow benchmark \citep{shadden05,brunton10}.
Coming back to the solar problem, we also 
found that the differences between the 2D plane results and the full
spherical analysis always remained relatively small. This is mostly
because the limb regions, where projection effects become very
important  and the surface velocity field itself can not be determined
accurately, were mostly ignored throughout the analysis. The spherical
and cartesian algorithms give almost identical results close to disc
centre.

\subsubsection{Imaging Lagrangian coherent structures}
Lagrangian coherent structures (LCS) are
simply defined as the ridges of a FTLE field $\sigma^T(\vec{x})$, and
as such follow directly from the computation of FTLEs
\citep{shadden05}. In 2D, they are most easily
understood as one-dimensional invariant manifolds acting as transport
barriers (although this is not strictly true, see the above paper for
a more accurate description), from which particles either diverge (repulsive
LCS), or to which they accumulate (attractive LCS). For a given flow, 
attractive LCS can be imaged using a (positive) FTLE field calculated from the
backward-in-time integration of tracers trajectories in that flow
(since maximal divergence for negative times implies maximal
convergence for positive times), while repulsive LCS are imaged
based on the (positive) FTLE field calculated from forward-in-time 
trajectories. We use both in what follows.

\section{FTLEs and Lagrangian Coherent Structures\label{LCSanalysis}}
In this Section, we describe global full-disc maps 
of Finite Time Lyapunov Exponents associated with large-scale 
solar surface convection flows, and further document their main 
properties.
\subsection{Global and local FTLEs and LCS maps}
\begin{figure*}
\includegraphics[width=\textwidth]{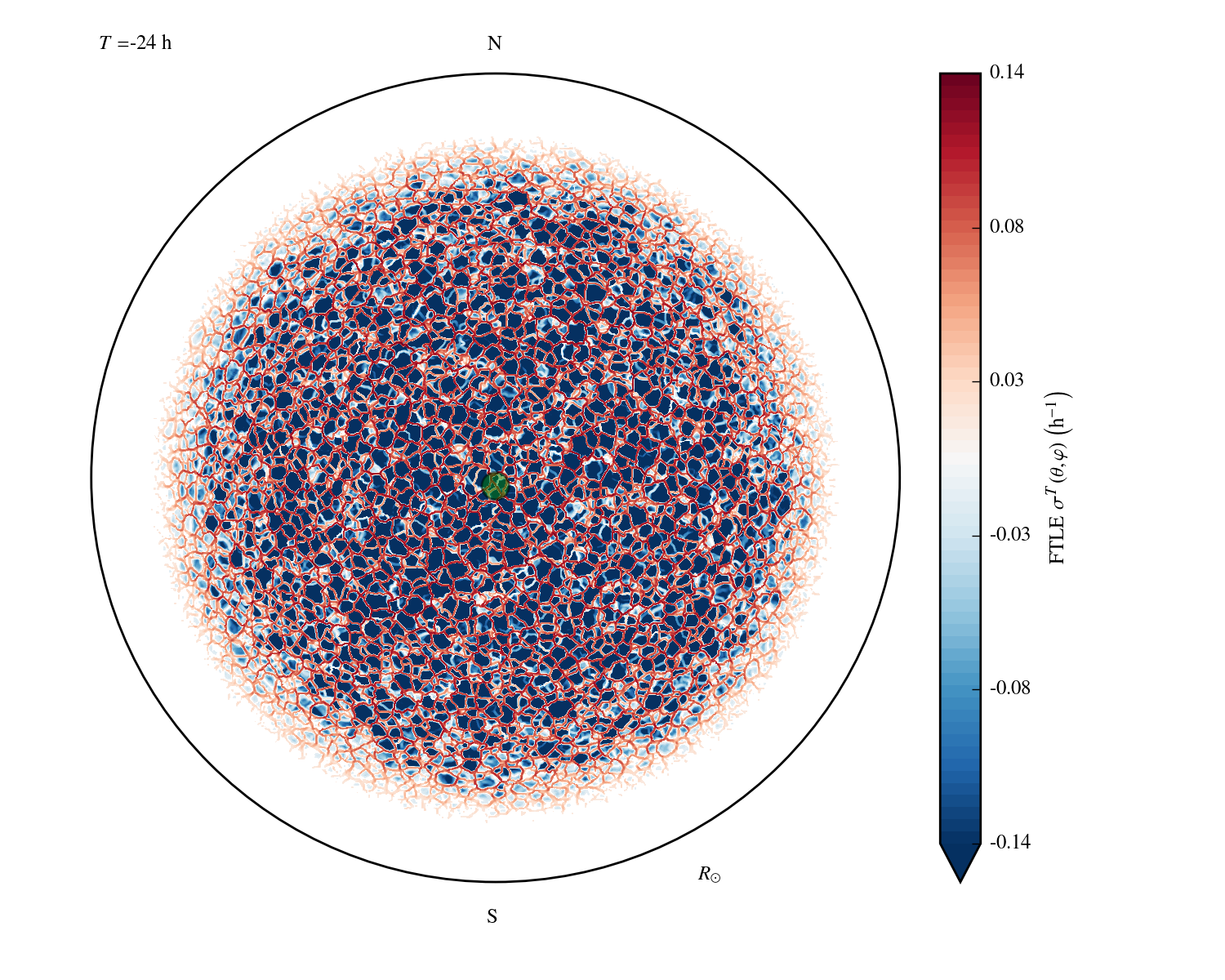}
\caption{\label{FTLEglobalfig} Global distribution of 24~h, backwards-in-time
  FTLE of solar surface flows (in inverse hour units) computed 
  using photospheric velocity field maps of 29 November, 2018, up to $60\degree$ 
from the disc centre. The green circle corresponds to a typical 30~Mm supergranule diameter.}
\end{figure*}

As an introduction, we first present in \fig{FTLEglobalfig} 
the global distribution of 24~h-negative
time FTLEs of solar surface flows (in inverse hour units) derived
from photospheric velocity field maps obtained on 29 November, 2018,
using the technique described in Sect.~\ref{fieldanalysis} and in Appendix.
This visualization reveals an (horizontally) isotropic tiling of the entire surface 
by structures/cells delimited by FTLE maxima, of size comparable to that of supergranules.
As explained earlier, the use of negative-time FTLEs outlines a network
of convergent/attractive (for positive times) spatial loci of passive 
tracers advected by the flow which, on this timescale, correspond to 
supergranule boundaries.

\begin{figure*}
\centering
 \includegraphics[width=\columnwidth]{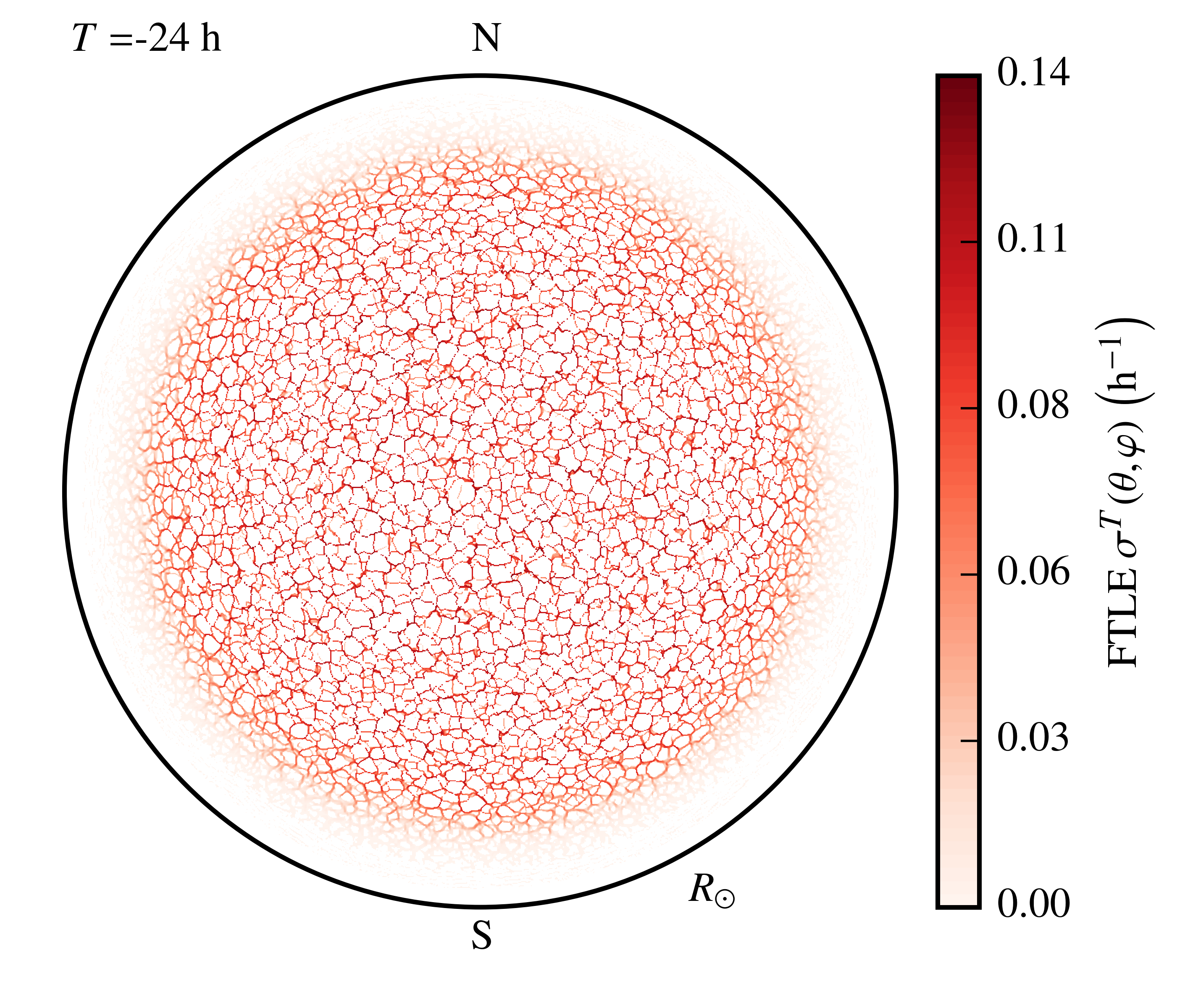}
 \includegraphics[width=\columnwidth]{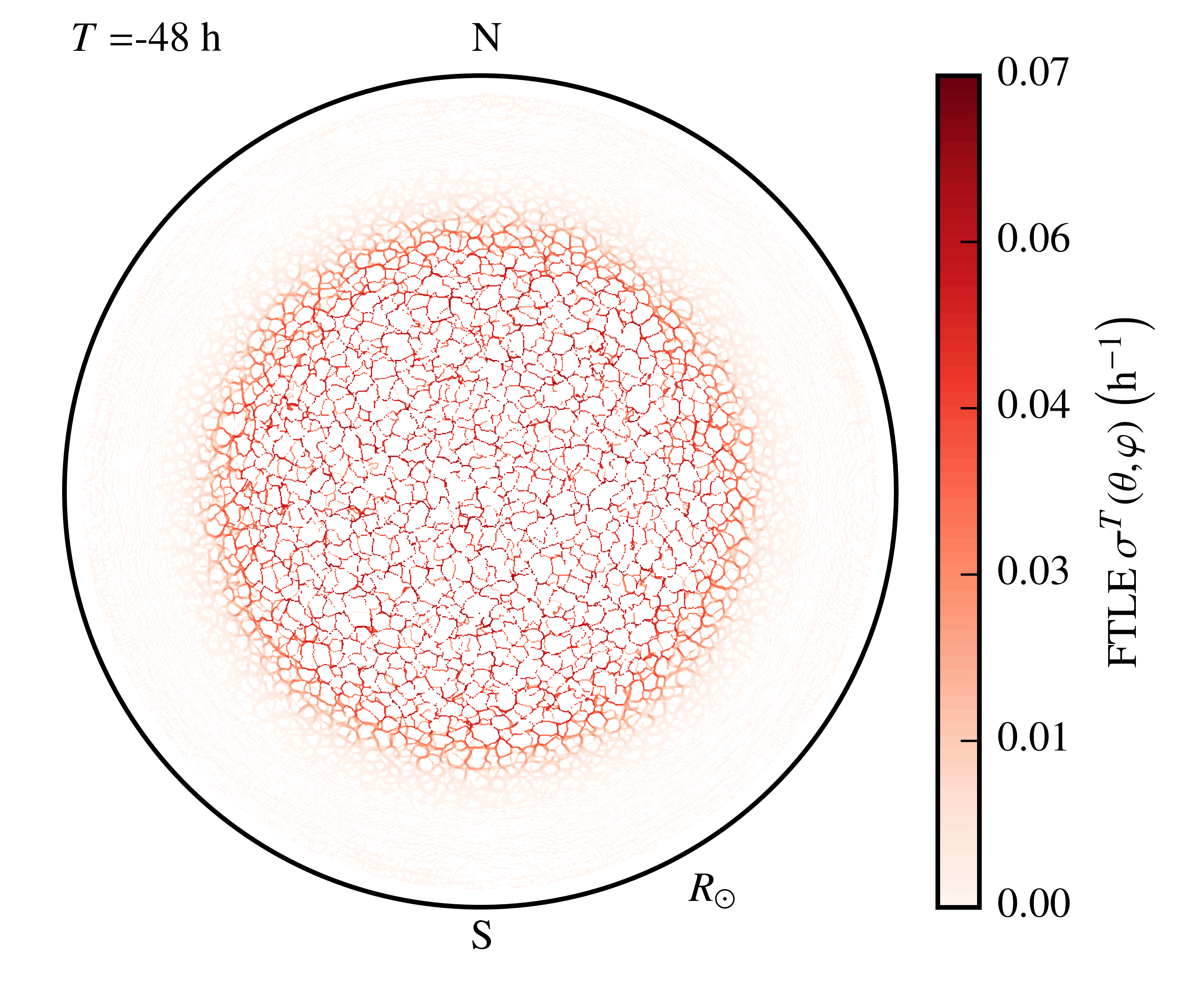}
\centering
\includegraphics[width=\columnwidth]{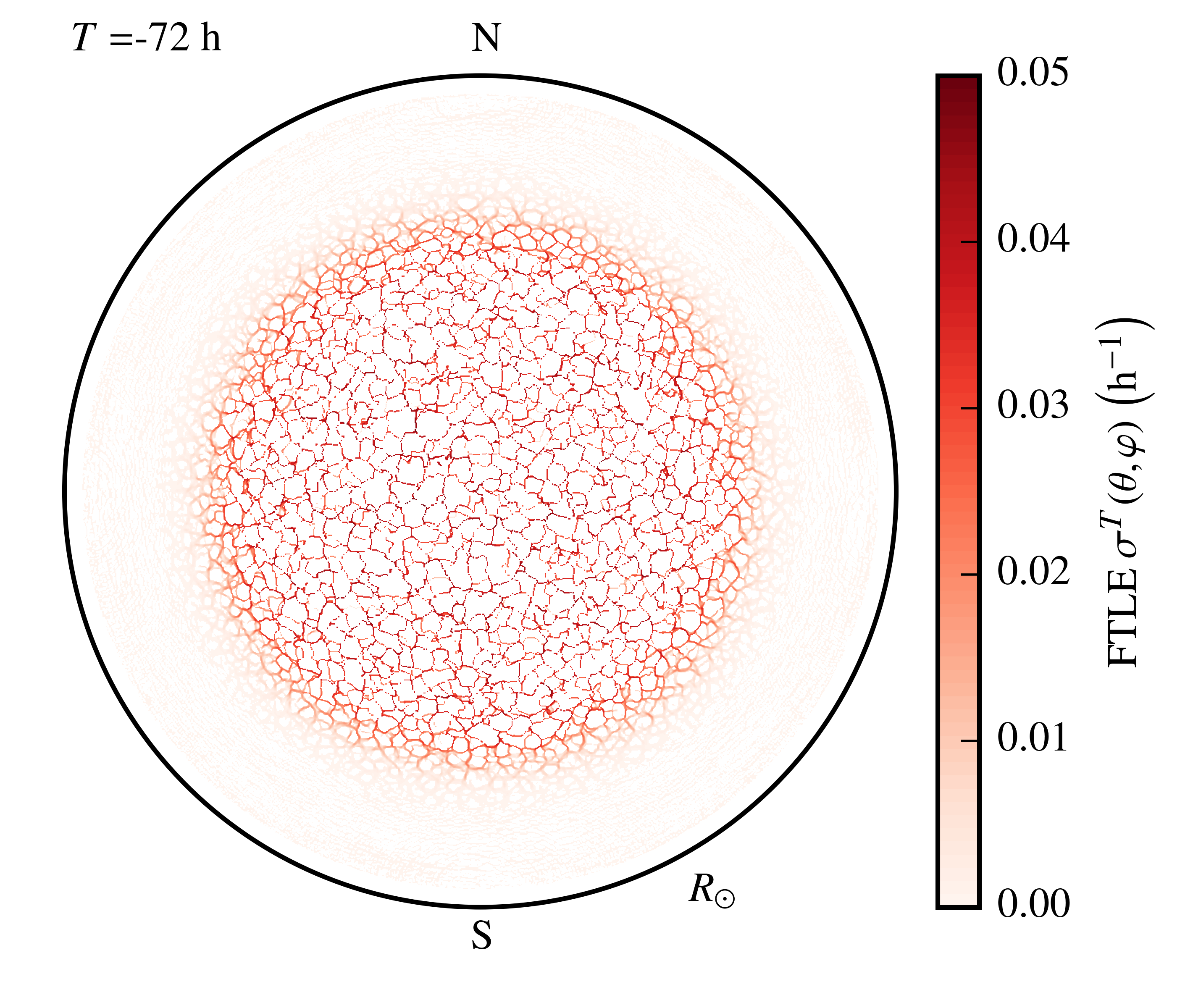}
\includegraphics[width=\columnwidth]{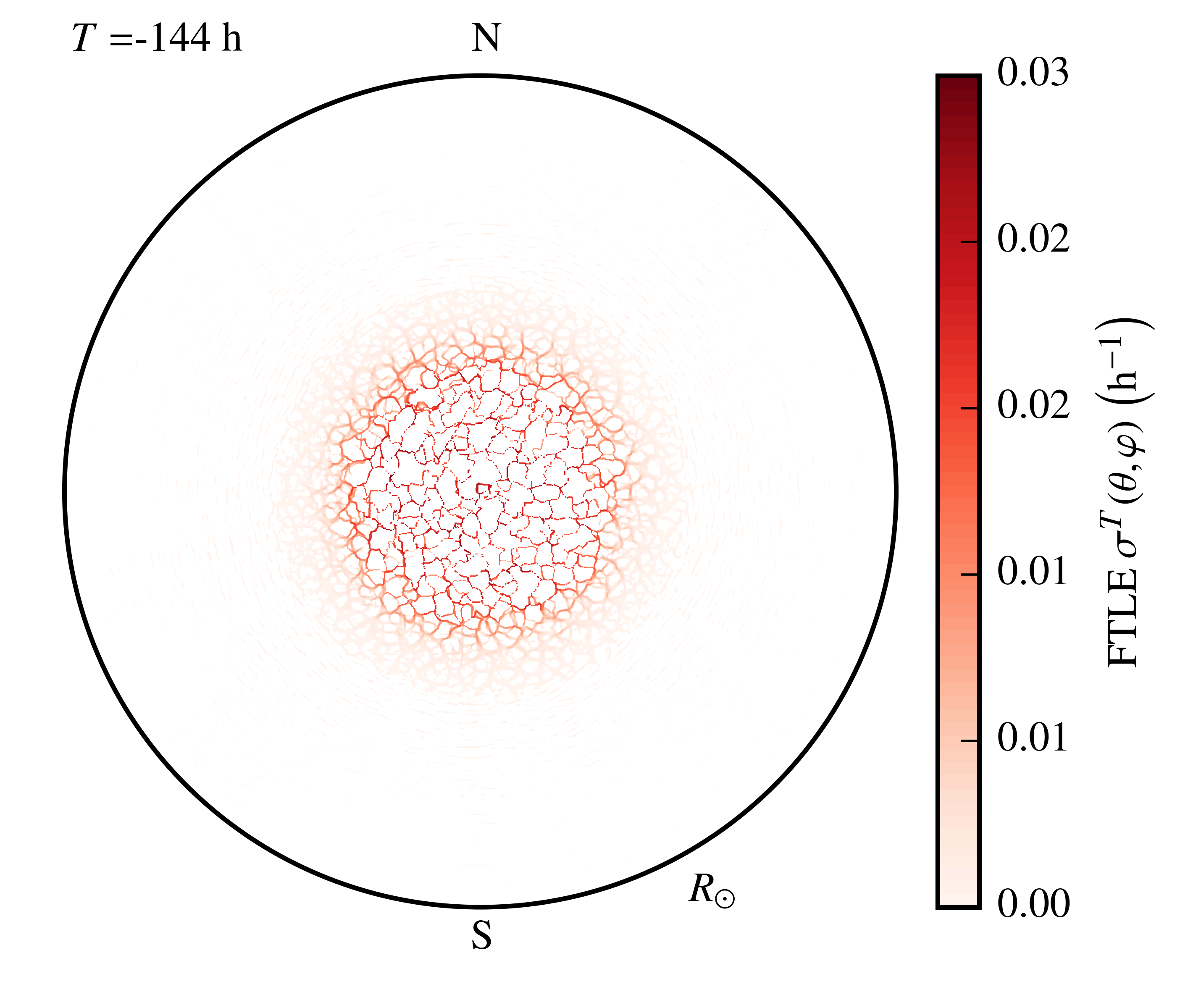}
\caption{\label{LCSfig}From left to right and top to bottom: global solar-disc maps of 
  attractive Lagrangian coherent structures imaged as ridges of maxima of the backwards-in-time 
  FTLE field, for different integration times of 1, 2, 3 and 6 days respectively. 
  Increasingly small apodising windows are used for longer time integrations (from $60\degree$ opening
  for $T=24~$h to $23.5\degree$ opening for $T=144$~h) to avoid contamination by imprecise velocity 
  measurements at the limb (see discussion in Sect.~2).}
\end{figure*}

This network can be further highlighted by looking at the ridges of the FTLE 
field, or by emphasizing the maxima of the field using an appropriate colormap. 
This results in maps of Lagrangian Coherent Structures (LCS) acting as a complex 
network of transport barriers on the timescale of consideration. In \fig{LCSfig}, 
we show such global maps for FTLE fields computed backward-in-time over 
different integration times, up to six days.  This is, to the best of our 
knowledge, the first calculation of this type over such long integration times
and large areas. The structures are remarkably long-lived, and simple visual inspection 
reveals their tendency  to moderately expand in size as a function of time. 
We characterise this effect quantitatively in Sect.~\ref{FTLEproperties}.

\subsection{LCS and the distribution of magnetic fields}
\begin{figure*}
\centering
\includegraphics[width=0.98\columnwidth]{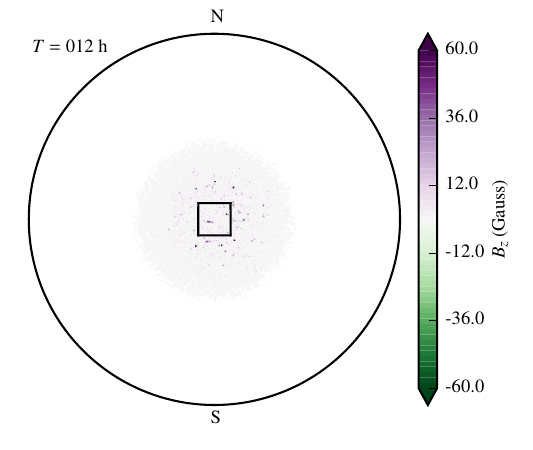}
\includegraphics[width=0.98\columnwidth]{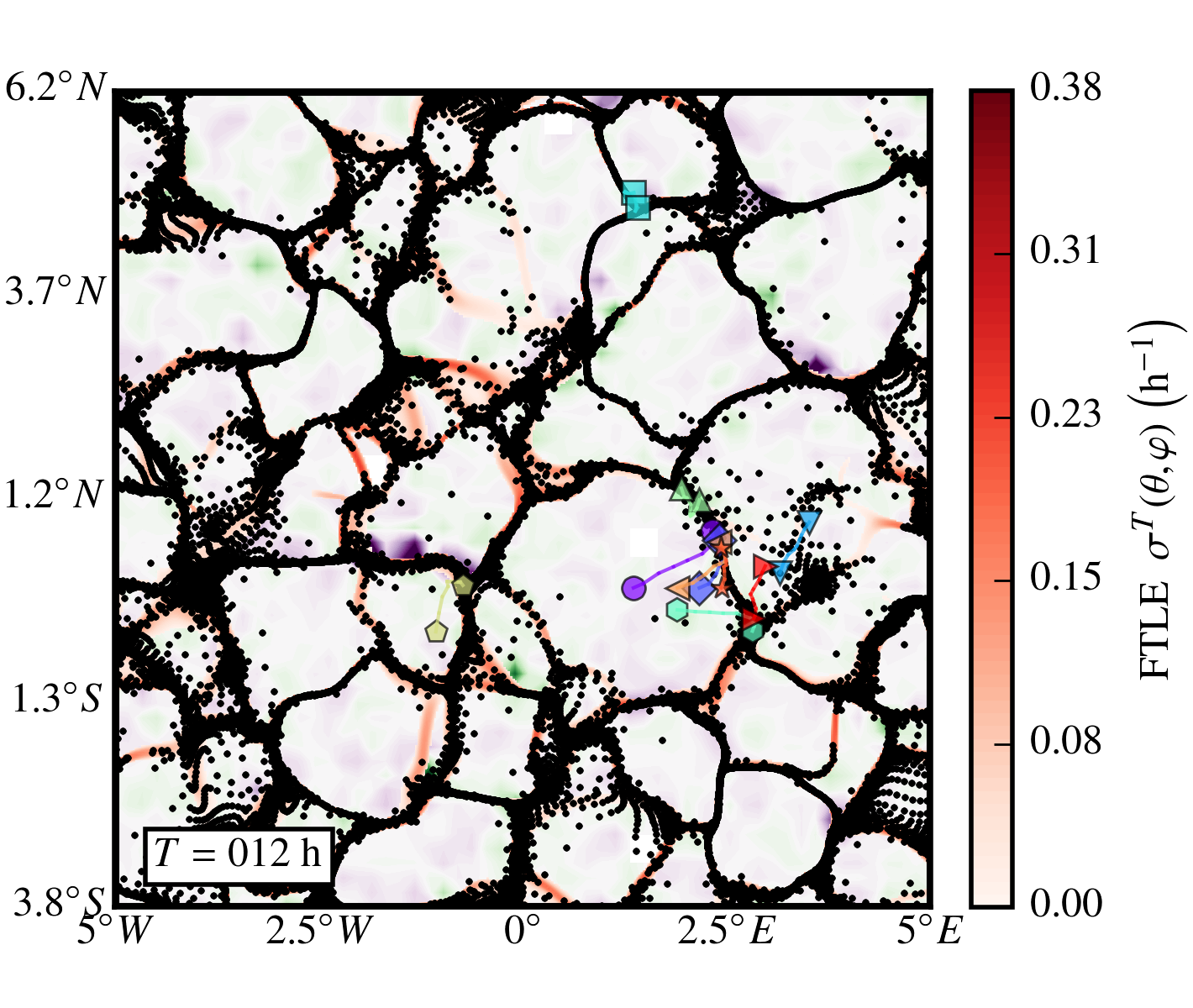}
\caption{\label{LCSlocalfig} Left: line-of-sight magnetic field $B_z$
over the central $23.5\degree$ region, imaged with SDO/HMI 12~h after the beginning of the 
observation. Right: zoomed-in centre-disc region of $10\degree\times 10\degree$
(frame in left plot) showing attractive LCS calculated 
for 12~h negative integration time (ridges of the negative-time FTLE field, red colormap),
superimposed with passive tracer locations after 12~h positive integration time (black dots), 
and corresponding local $B_z$ magnetograms (green/purple colormap). The coloured markers and lines
tag 12~h (positive time) trajectories of a few selected tracers, see discussion in 
Sect.~\ref{transport} and \fig{diffusionfig} below (visualization continued for longer 
times in \fig{LCSlocalfig2}).}
\end{figure*}

\Figs{LCSlocalfig}{LCSlocalfig2} show zooms of such maps on a region of $122\times122$~Mm$^2$ ($10\degree\times 10\degree$)
at the (derotated) disc centre, overimposed with SDO/HMI magnetograms. These zoomed maps do not only enable us to 
better appreciate the mesmerizing 
structure of these patterns, they also show how much their fine-scale structure correlates with the magnetic
network at the surface. As discussed by \cite{yeates12}, on timescales of the order of the supergranulation 
timescale, LCS should be associated with regions of magnetic field accumulations. This is exactly what we observe here
too: small-scale photospheric magnetic field concentrations appear to correlate extremely well with LCS derived 
from passive tracers. The results above (\figs{FTLEglobalfig}{LCSfig}) therefore offer a striking new global-scale 
perspective on this phenomenon. Finally, \figs{LCSlocalfig}{LCSlocalfig2} show the trajectories of a few selected tracers
with remarkable excursions on the timescales of consideration. We will discuss these in more detail in Sect.~\ref{transport}
in connection with magnetic dynamics and transport.

\begin{figure*}
\centering
\includegraphics[width=0.98\columnwidth]{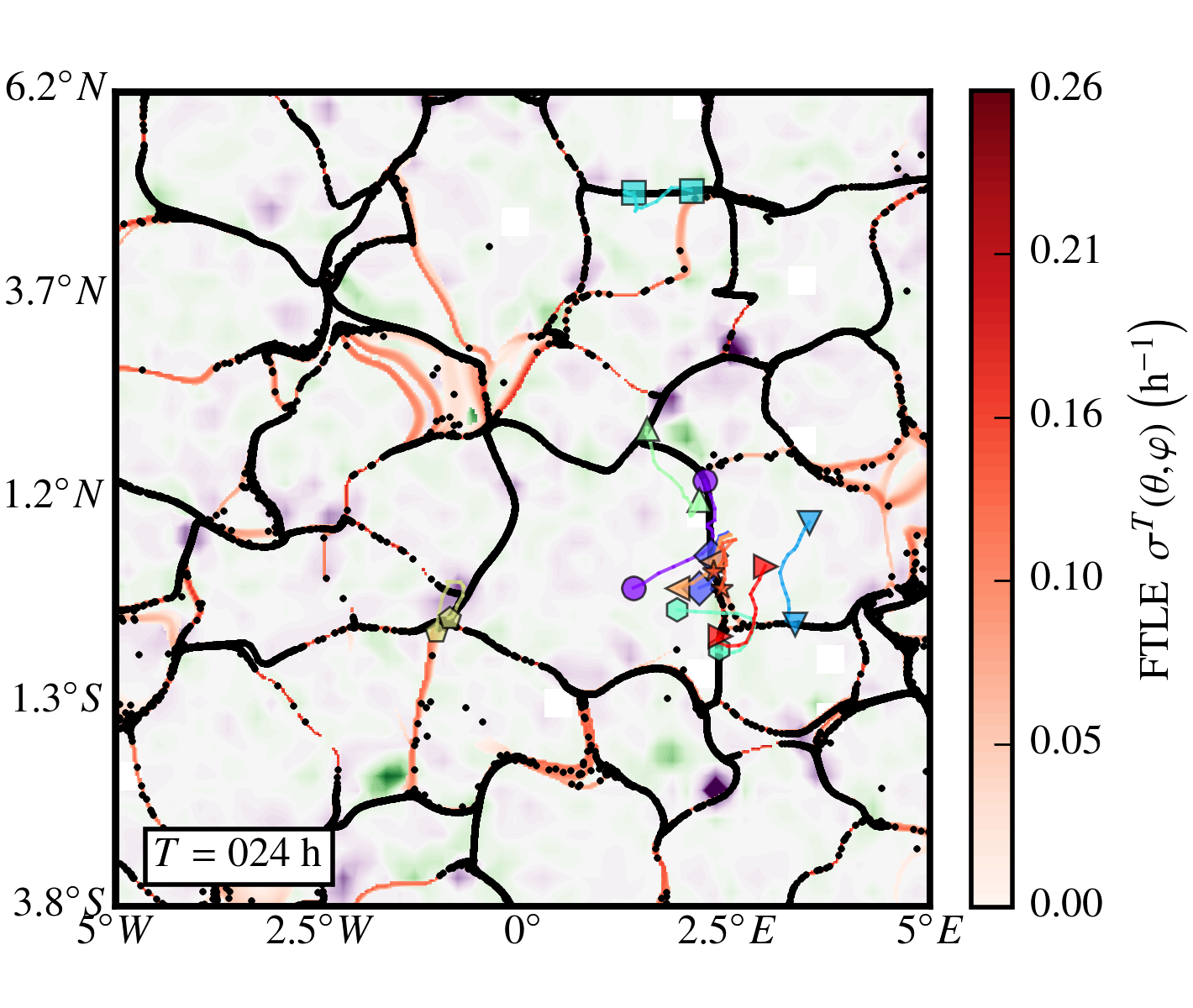}
\includegraphics[width=0.98\columnwidth]{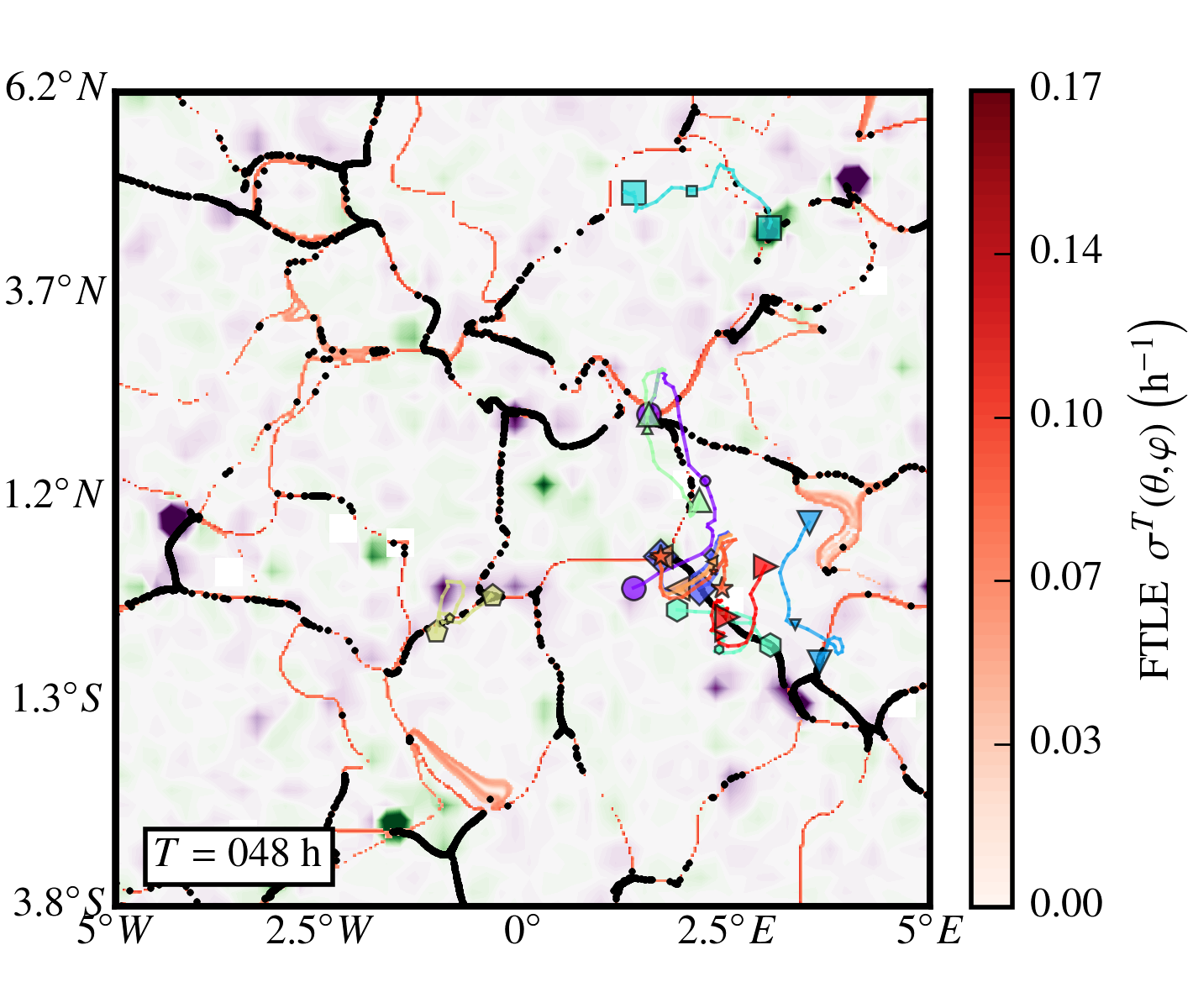}
\centering
\includegraphics[width=0.98\columnwidth]{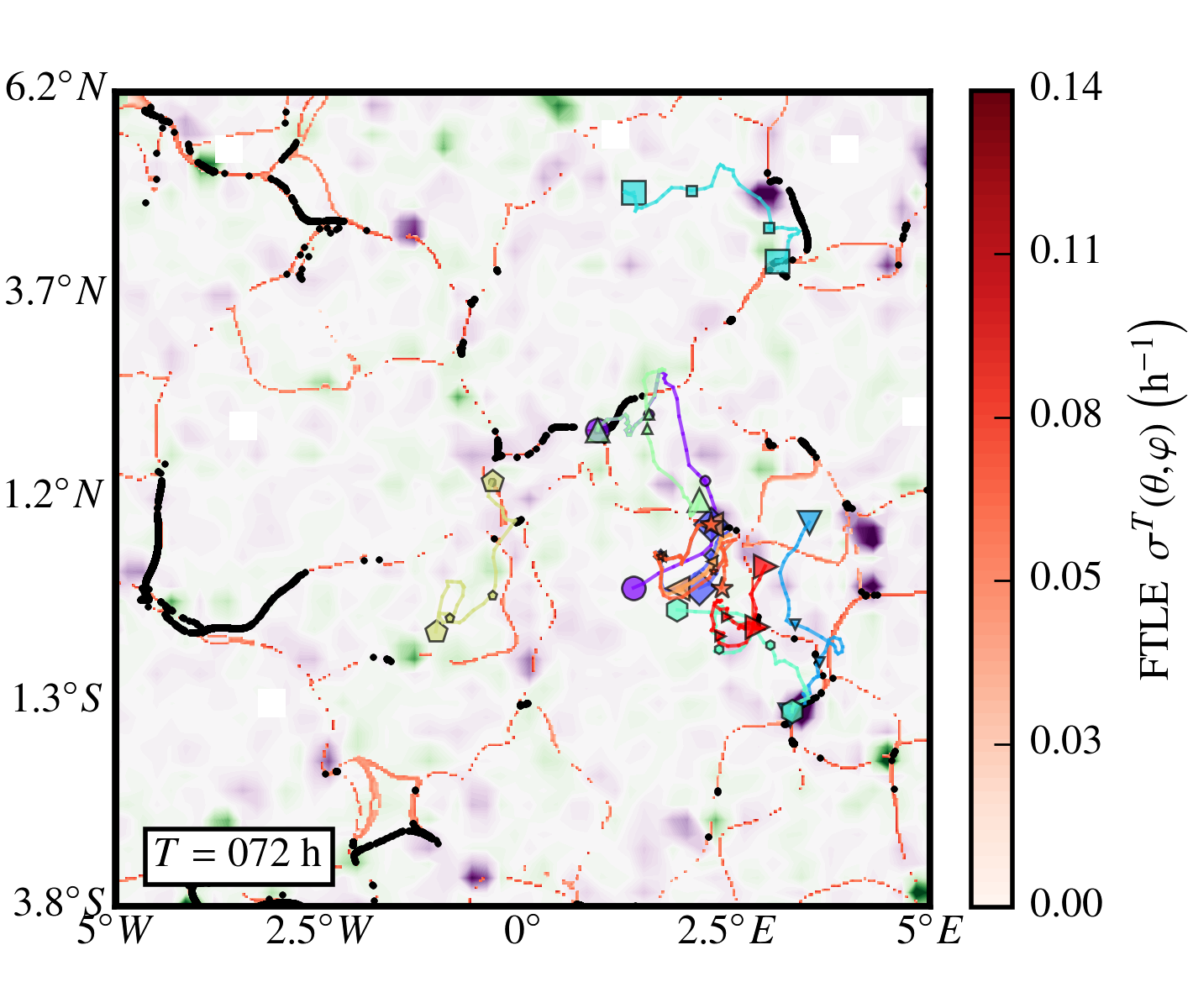}
\includegraphics[width=0.98\columnwidth]{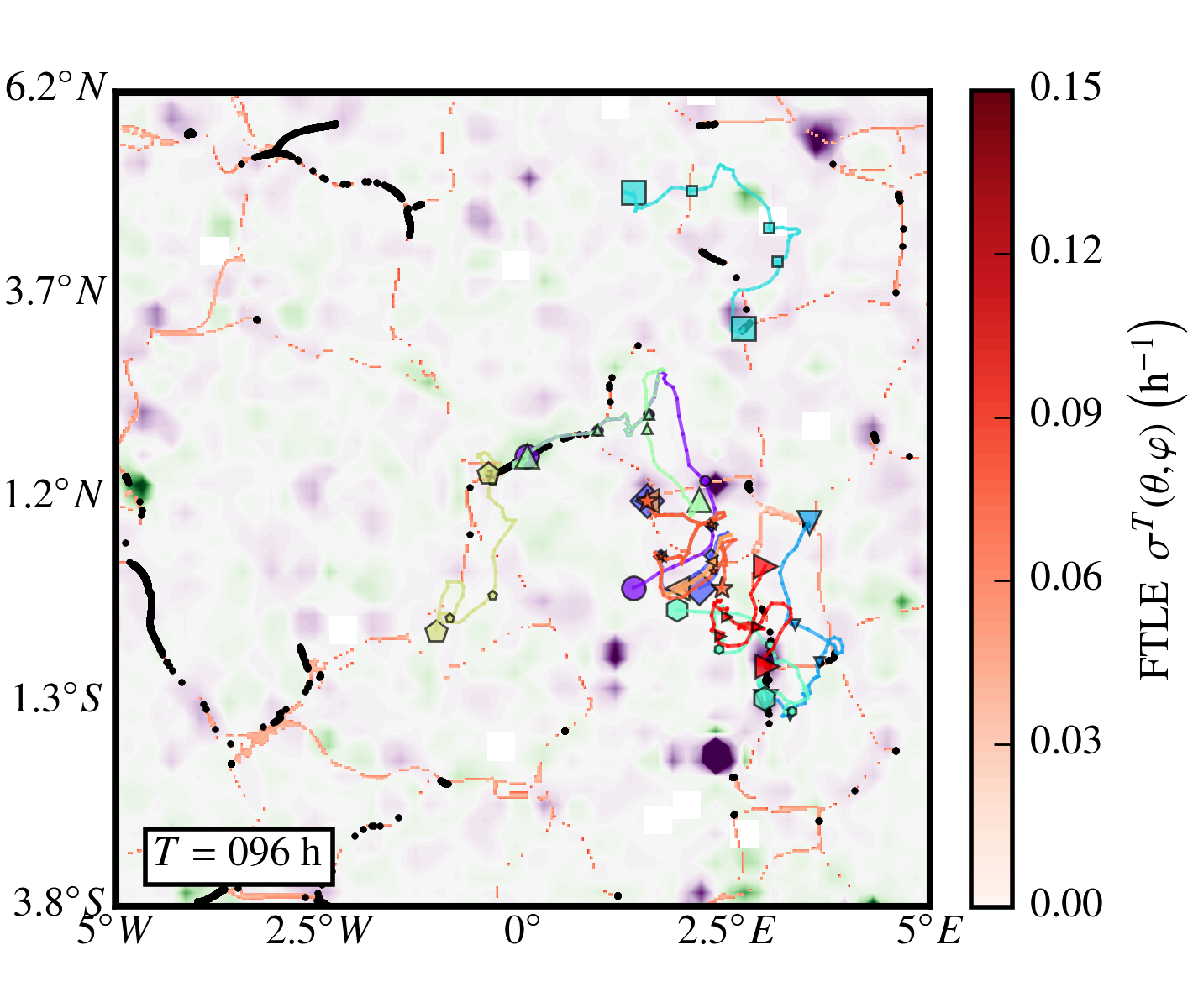}
\centering
\includegraphics[width=\columnwidth]{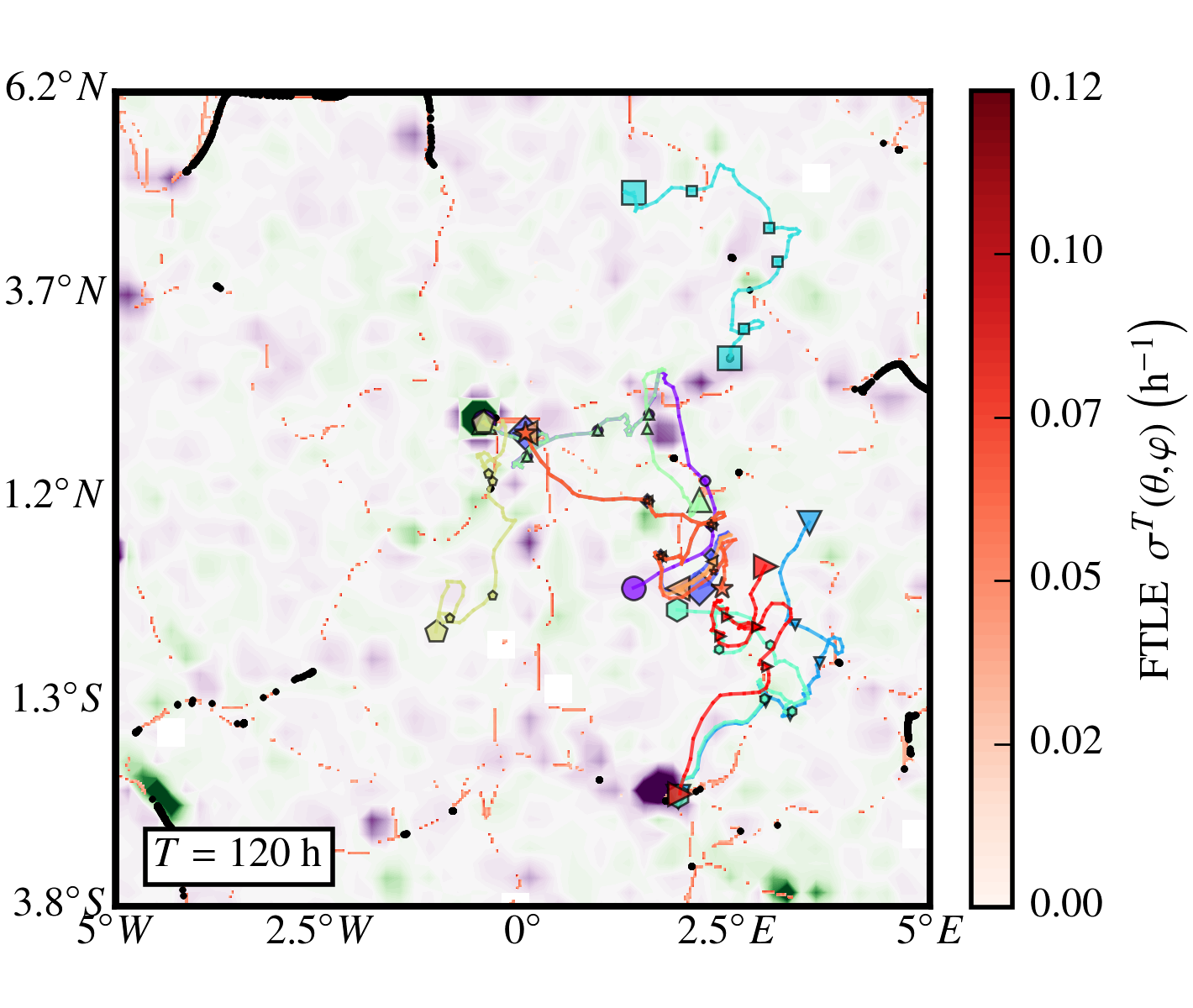}
\includegraphics[width=\columnwidth]{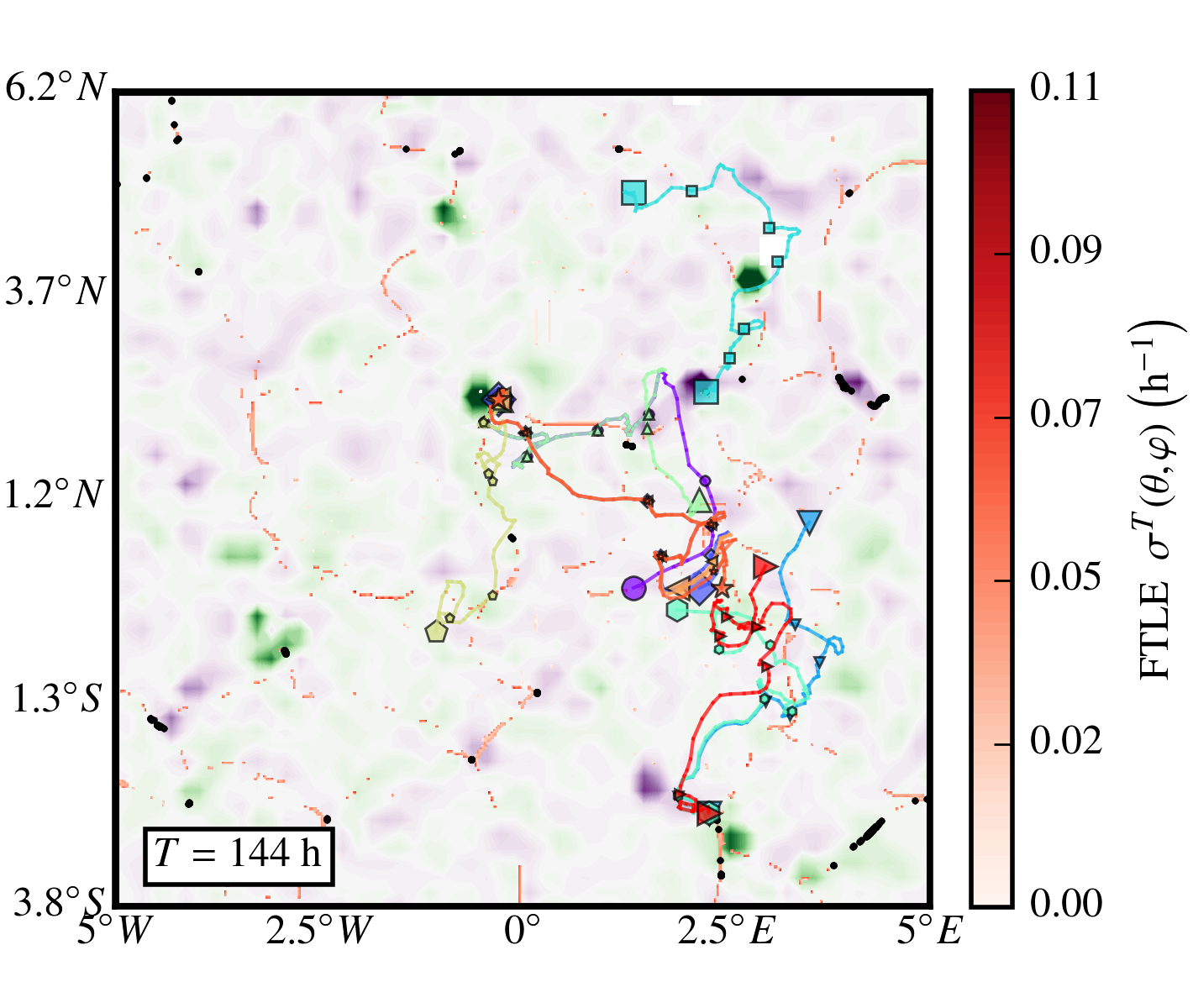}
\caption{\label{LCSlocalfig2} Continuation of \fig{LCSlocalfig}.
From top left to bottom right: zoomed-in region 
(frame in \fig{LCSlocalfig}, left) showing attractive
LCS calculated for successive negative tracer 
integration times (red colormap), superimposed with passive tracer 
locations (black dots) and corresponding local $B_z$ magnetograms 
(same green/purple colormap as in \fig{LCSlocalfig}) 
at corresponding positive integration times.
As mentioned in Sect.~\ref{fieldanalysis}, tracers accumulate 
in attractive LCS. Throughout the sequence, LCS and tracer concentrations correlate 
well with the magnetic network and bright points, respectively.
The coloured markers and lines tag (positive-time) trajectories of a few
selected tracers: large symbols mark the initial position 
and current position for the time of integration of each plot, and small symbols 
the position for every intermediate 24~h (see discussion in Sect.~\ref{transport} 
and \fig{diffusionfig}).}
\end{figure*}

\subsection{Spectrum, scales and strength of the FTLE field\label{FTLEproperties}}
To be more quantitative, for any FTLE map such as shown in \fig{FTLEglobalfig}, 
we compute a typical peak scale of the distribution of FTLEs, corresponding to the 
wavelength of the LCS/transport barriers, as the integral scale of the FTLE field. 
To do this, we simply treat each FTLE field as a scalar field $\sigma^T(\theta,\varphi)$ 
over the sphere apodised by the visibility window, take its harmonic transform, 
compute the associated spherical harmonics spectrum $E_\sigma(\ell,T)$, and calculate 
the integral scale of the field, defined as
\begin{equation}
\label{Lsigmaeq}
L_\sigma(T)= \frac{\displaystyle{\sum_{\ell>50} \f{2\pi R_\sun}{\ell}\, E_\sigma(\ell,T)}}{\displaystyle{\sum_{\ell>50} E_\sigma(\ell,T)}}~.
\end{equation}
This formula simply weighs each scale by its corresponding energetic spectral content, 
providing a weighted mean giving more weight to the scales at which the distribution of $\sigma^T$ 
contains the most energy.  The lower bound in the sum avoids contamination by the large-scale energetic 
content of the apodising window, with which the true field is convoluted in spectral space. 
The numerical technique used for the spectral harmonic decomposition of our data fields over 
apodised fields of views is described in detail in R17, where ample use of it was made to 
characterise the structure and scale of the turbulent flow itself.  Here, we simply apply 
the same tool to FTLE fields. 

\begin{figure}
\includegraphics[width=\columnwidth]{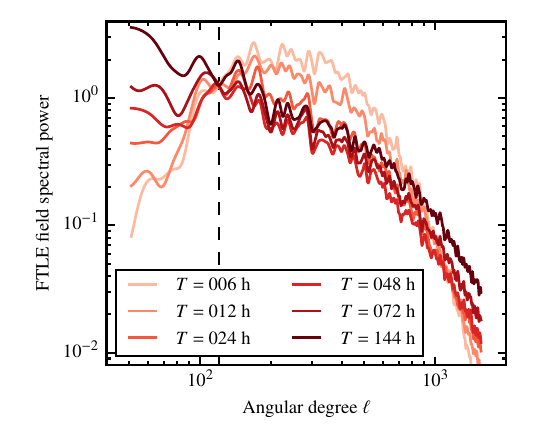}
\caption{\label{FTLEspec}Spherical harmonics spectra of the FTLE field for different target times. 
The spectra are normalized to one at the standard supergranulation angular degree scale $\ell=120$ 
(dashed vertical line) to highlight the pivot of the spectrum towards larger physical angular scales 
on longer times.}
\end{figure}

\fig{FTLEspec} shows the spherical harmonics spectra $E_\sigma(\ell,T)$ of FTLE fields for different 
target integration times, shown earlier in \fig{LCSfig}. As LCS form and evolve, their energy content 
and peak scale shifts towards larger scales, and the spectrum becomes steeper. By $T=24~$h, 
the slope of the spectrum becomes stationary at intermediate scales, but the peak scale continues 
to increase monotonically on larger times.

\begin{figure}
\includegraphics[width=\columnwidth]{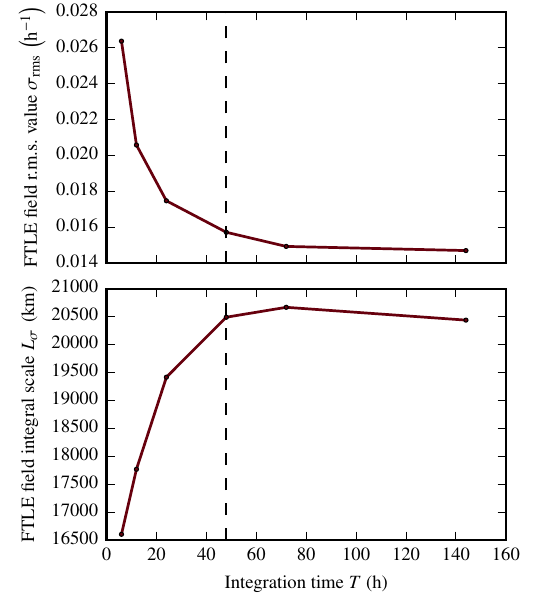}
\caption{\label{FTLEscalefig}Change as a function of the FTLE target integration time 
of the r.m.s. value and integral scale of FTLE distributions, derived from spherical harmonics spectra 
using \equ{Lsigmaeq}. The dashed vertical line shows $\tau_\mathrm{SG}=48~$h, the typical correlation time of 
supergranulation.}
\end{figure}

The corresponding evolution of $L_\sigma$ is shown in \fig{FTLEscalefig} as a function of 
the target time of integration of passive tracers. As this time increases, $L_\sigma$ increases 
from 16~Mm at $T=6$~h to 21~Mm at $T=72$~h, and subsequently saturates. Thus, LCS, while acting as 
transport barriers, are "not set in stone" in this kind of time-dependent flow unconstrained by 
horizontal boundaries\footnote{In marine dynamics for instance, LCS can be laminar patterns whose geometry is 
shaped by coastal/bay topography, and they can be extremely stable over time, leading for example to 
accumulation of pollutants in particular areas, see e.g. \cite{lekien05}.}. 
Instead, they themselves appear to be highly dynamical structures that change in time as 
the flow evolves. 

\Fig{FTLEscalefig} also shows how the r.m.s. value of the FTLE field $\sigma^T$ evolves with the integration time.
This quantity indicates how particles diverge as a function of time. A fast initial divergence over a timescale
of a few hours is initially observed ($\sigma_\mathrm{rms}\sim 0.025~\mathrm{h}^{-1}$, with peak values 
$\sigma_\mathrm{max} \sim 0.14~\mathrm{h}^{-1}$ for $T=24$~h corresponding to a shortest divergence timescale 
of just 8~h, see \fig{FTLEglobalfig}). This shorter timescale roughly corresponds to the time it takes for tracers 
to be swept by supergranulation-scale flows from a supergranulation cell centre to its periphery. On longer timescales,
however, particles tend to get trapped/blocked in the LCS transport barriers associated with supergranule boundaries.
As a result, the rate of relative divergence of their trajectories significantly decreases, as visible 
in \fig{FTLEscalefig}, and appears to saturate at $\sigma_\mathrm{rms}\sim 0.015~\mathrm{h}^{-1}$.

Using the long-time asymptoting values reached by $L_\sigma$ and $\sigma_\mathrm{rms}$ in \fig{FTLEscalefig},
we can infer a first rough dimensional turbulent diffusion coefficient estimate, namely
\begin{equation}
\label{Dsigmaeq}
D_\sigma=\f{\sigma_\mathrm{rms} L^2_\sigma}{4}\simeq 400~\mathrm{km}^2\,\mathrm{s}^{-1} = 4\times 10^8~\mathrm{m}^2\,\mathrm{s}^{-1}~.
\end{equation}
While this simple result should be treated with caution and should only be seen as an order of magnitude estimate,
we point out that it provides an interesting connection between LCS, FTLEs and large-scale 
transport, which we study more quantitatively in the next section.

\section{Statistical characterisation of turbulent transport\label{transport}}
The previous section aimed at providing a phenomenological perspective on large-scale transport
in the quiet photosphere,  and it notably enabled us to pinpoint a global network of Lagrangian 
coherent structures associated with supergranulation-scale convection as a key physical pattern
regulating horizontal turbulent diffusion in this convective fluid system.  We now proceed 
to analyse the horizontal transport process on timescales of up to six days from a more classical
statistical perspective, and subsequently attempt to interpret the results through the 
phenomenology outlined previously.

\subsection{Statistics of tracer trajectories\label{stattracers}}
To quantify turbulent transport, we focus on the central $23.5\degree$ region, whose velocity
field can be reliably inferred throughout our 144~h sequence.
We calculate the trajectories of $1024\times 1024$
tracers initially placed on a cartesian grid at the centre of this region, with a 
resolution of $135~\mathrm{km}$. We first calculate the ensemble average (denoted by brackets) of the squared 
distance $d^2_{i,j}(T)=|\vec{x}_{i,j}(T)-\vec{x}_{i,j}(0)|^2$ travelled by each $(i,j)$ tracer initially 
placed on this cartesian grid, as a function of the integration time. On general grounds, we 
expect 
\begin{equation}
\label{distanceeq}
\langle d^2\rangle(T)= c T^{\gamma}~,
\end{equation}
with $\gamma=1$ corresponding to a random walk diffusion regime, $\gamma>1$ to an anomalous 
superdiffusion regime, and $\gamma<1$ to an anomalous subdiffusion regime. 
Using power-law fits to \equ{distanceeq}, we can then calculate a horizontal 2D-turbulence
diffusion coefficient using a formula frequently used in the solar physics context 
\citep[][see e.g. \cite{hagenaar99,abramenko11}]{monin71},
\begin{equation}
\label{diffcoeffformula}
D=\f{1}{4}\frac{\mathrm{d}\,\langle d^2\rangle}{\mathrm{d}\,T}=\frac{\gamma\, c}{4}\, T^{\gamma-1}~.
\end{equation}
Note that this transport coefficient usually bears a residual
dependence on $T$ if $\gamma\neq 1$, and that this expression also motivates \textit{a posteriori} 
the specific expression used in \equ{Dsigmaeq} to estimate turbulent diffusion on the basis 
of the characteristic rate of tracers divergence and spatial scales obtained from our FTLE analysis.

\begin{figure}
\includegraphics[width=\columnwidth]{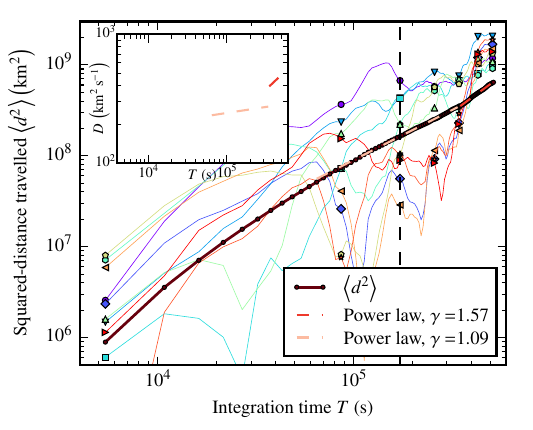}
\caption{\label{diffusionfig}{Ensemble-averaged $\langle d^2\rangle$ of the 
quadratic distance travelled by tracers as a function of time (full dark red line), 
power law fits (lighter red and cream lines), and (inset) corresponding diffusion 
coefficients as determined by \equ{diffcoeffformula}. The thin coloured lines with markers 
every 24~h show the history of $d^2$ for individual tracers with 
larger-than-average excursions on long times. 
The surface trajectories of these tracers are shown with the same markers and 
colours in \figs{LCSlocalfig}{LCSlocalfig2}.}}
\end{figure}

\begin{figure}
\includegraphics[width=\columnwidth]{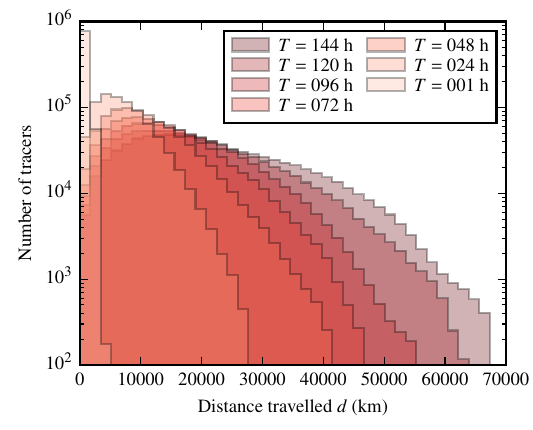}
\caption{\label{diffhistogramfig}{Time evolution of the histogram distribution 
of distances travelled by tracers. The tracers showcased by markers 
in \figs{LCSlocalfig}{LCSlocalfig2} and \fig{diffusionfig} 
all belong to the tail of the distribution at $T=144$~h.}}
\end{figure}

We plot $\langle d^2\rangle(T)$ and $D$ for our collection of advected tracers in \fig{diffusionfig}.
We obtain a reasonable power-law fit with $\gamma=1.09$ for intermediate times comparable to 
$\tau_\mathrm{SG}=48~$h, the typical correlation time of supergranulation. This corresponds to an almost-diffusive
behaviour with a turbulent diffusion coefficient in the range $D=200-300~\mathrm{km}^2\,\mathrm{s}^{-1}$ (inset).
A histogram of the distribution of distances travelled by tracers (\fig{diffhistogramfig}) further illustrates
the spread in distances travelled as a function of time, consistent with a diffusive process.

On the longest times probed though, we find in \fig{diffusionfig} a slight increase in 
transport ($\gamma=1.57$), corresponding to an enhanced diffusion 
coefficient $D\simeq 400~\mathrm{km}^2\,\mathrm{s}^{-1}$. To understand the origins of this trend,
we isolated a collection of tracers with larger-than-average excursions $d$ from their initial positions.
We overplot $d(T)$ for ten of these tracers in the figure using thin color lines and symbols,
and further trace their trajectories at the solar surface in \figs{LCSlocalfig}{LCSlocalfig2}. A 
careful examination of these figures reveals that most of these tracers 
experience secondary transport kicks at a time between 24~h and 72~h comparable to $\tau_\mathrm{SG}$. 
These kicks appear to be associated with the regeneration of the supergranulation flow 
and the emergence of new "explosive granules" \citep{roudier16} that structure supergranules,
leading to a transient, more ballistic-like energetic superdiffusive behaviour.  Since the average 
statistical trend at the largest times probed here is dominated by the small collection of tracers 
in this "non-thermal" tail of the distribution shown in \fig{diffhistogramfig}, we argue, consistent
with the standard picture of a random walk, that such individual kicks are simply
random-walk events associated with the regeneration of the supergranulation flow,
and that the accumulation of such events on timescales much larger than $\tau_\mathrm{SG}$  
would likely result on long times in an average diffusive-like behaviour. Accordingly, we conjecture
that the statistics of $\langle d^2\rangle$ should eventually settle in such a large-scale turbulent 
diffusion regime if we could track tracers trajectories for even longer times than we did here 
to reach a better timescale separation with the typical correlation time of the flow.
Despite our best efforts to maximize the consecutive time of observation, reaching this regime 
unfortunately currently remains impossible, due to the observing limitations stressed in 
the Introduction and in Sect.~\ref{datalimitation}.

With respect to this point, close examination of \figs{LCSlocalfig}{LCSlocalfig2} 
also reveals that some magnetic concentrations
at the surface are subject to the exact same dynamics as these outlier tracers. For instance, 
an intense violet magnetic concentration located at $(2\degree E, 2.3\degree S)$ at $T=96$~h 
gets pushed further south in exactly the same way as the tracers tagged with red and cyan triangles 
between $T=96$~h and $T=120$~h, at which time the position of the two tracers almost coincides
with that of the magnetic element. The same remark applies at $T=120~$h to the green magnetic concentration
located at $(0.5\degree W, 2.2\degree N)$, and to the tracer tagged by a yellow pentagon. 
This provides further confirmation that the dynamics of tracers and magnetic fields are 
strongly correlated, and that both are subject to the same kind of average transport and 
individual transport events/kicks.

\subsection{Ink spot experiment}
To complement the previous results, and in the spirit of approaching the problem from a
large-scale statistical point of view, we finally present a slightly different 
characterisation of transport by photospheric flows. The idea is to mimic a classical 
ink spot molecular diffusion experiment in the lab, whereby a circular ink spot is 
carefully released in water, and its radius subsequently expands diffusively. To do this, 
we delimit a circular patch of tracers in the initially cartesian grid of tracers
(as in Sect.~\ref{LCSanalysis}, we use a large tracers grid covering the entire solar surface here).
We then measure the evolution of their average squared distance from the centre of the spot, i.e.
\begin{equation}
\label{rspoteq}
R^2_\mathrm{spot}(T)=\f{1}{N}\sum_{i,j\,\in\,\mathrm{spot}}\left(x_{i,j}(T)-x_{\mathrm{spot},c}\right)^2+\left(y_{i,j}(T)-y_{\mathrm{spot},c}\right)^2~,
\end{equation}
where $(x_c,y_c)$ are the central coordinates of the spot on the CCD grid and $N$ is the number of 
tracers initially within a radius $R_0$ of the spot centre.  
For a diffusive process, a spot of initial radius $R_0$ will diffuse on a timescale $\tau_D=R_0^2/D$, 
where $D$ is the diffusion coefficient. For the experiment to be meaningful, we should pick $R_0$ small enough
that $\tau_D$ is smaller than, or at most of the order of the observation period, but also large enough
that the circular spot itself encompasses at least one typical flow structure, i.e. a supergranule.
Indeed, a very large spot would barely start to diffuse on the available time of observation, while a very
small one would not feel the statistical effect of turbulent kicks and would also not contain enough 
tracers to construct a meaningful statistics. In what follows, we chose $R_0=28.7$~Mm, which corresponds 
to an angle with respect to the Sun centre of $2.35\degree$, one-tenth of the opening of the 
apodising window used for the continuous six-day observation. Based on our earlier estimate for $D$, 
such a spot should diffuse on a typical timescale of a month, so we expect to be able to measure 
the beginning of the diffusion process using our six-day sequence. 

A spot of initially this size contains $\sim 1380$ tracers for our grid of tracers  
and it samples the flow of only a handful of independent supergranules giving relatively
noisy results.
To improve the statistics, we therefore replicate the measurement in \equ{rspoteq} for as many spots as 
possible, packing the $23.5\degree$ observation window with non-overlapping circular spots whose centres are
contained initially within an opening $21.15\degree=23.5\degree-2.35\degree$ angle from the solar 
disc centre. For the chosen $R_0$, this gives us a statistics of 60 spots from which we 
can calculate $\overline{R^2_\mathrm{spot}}$, a spot-ensemble average of $R^2_\mathrm{spot}$. 
Because the diffusion process takes a few hours to develop, we parameterise the evolution as
\begin{equation}
\label{Rspotaverageeq}
\overline{R^2_\mathrm{spot}}(T)=R^2_\mathrm{spot}(0)\left(1+c T^\gamma\right)~.
\end{equation}

\begin{figure}
\includegraphics[width=\columnwidth]{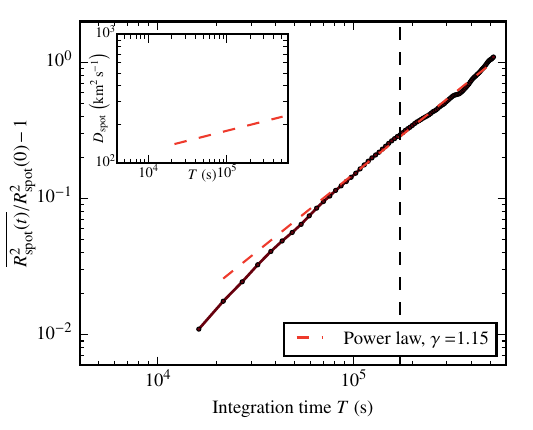}
\caption{\label{diffusionspotfig}Time evolution of the ensemble-average normalised 
radius of circular spots of tracers (full dark red line, see \equ{Rspotaverageeq}), 
and correspondonding power-law fit.}
\end{figure}

\begin{figure}
\includegraphics[width=\columnwidth]{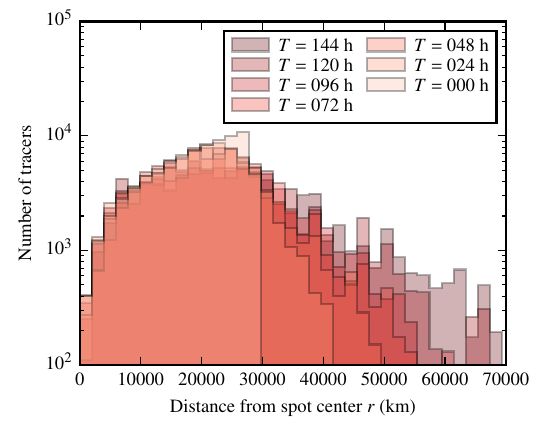}
\caption{\label{diffspothistogramfig}Time evolution of the histogram distribution 
of distances travelled by tracers initially enclosed within circular
spots of tracers of radius 30~000~km, with respect to their relative 
spot centres. Together with \fig{diffusionspotfig}, the plot highlights 
the turbulent diffusive spread of tracers.}
\end{figure}

In \fig{diffusionspotfig}, we plot $\overline{R^2_\mathrm{spot}}/R^2_\mathrm{spot}(0)-1$ as a function of 
the integration time of tracers, and a corresponding power law fit of the evolution for times longer 
than 24~h.  While the observed statistics is neither an exact power law, nor exactly 
diffusive, diffusion still appears to be a very reasonable first approximation overall, and
the scaling exponent $\gamma=1.15$ is very similar to $\gamma=1.09$ obtained with the first method 
used in Sect.~\ref{stattracers}.  As in \fig{diffusionfig}, we also observe that transport is slightly 
enhanced on late times, for the same reasons. \Fig{diffspothistogramfig} further shows the histogram distribution,
as a function of time, of the distance of all the tracers considered in the analysis from their relative 
spot centre. The results illustrate the diffusion-like spread of the averaged spot.

Using this analysis, we can now estimate a large-scale diffusion coefficient from the power 
law fit in \fig{diffusionspotfig}. For consistency and comparison with the method 
used in Sect.~\ref{stattracers}, we use again \equ{diffcoeffformula} for this purpose, 
applied here to $\overline{R^2_\mathrm{spot}}/R^2_0-1$ instead of $\langle d^2\rangle$. 
The results, shown in the inset of \fig{diffusionspotfig}, 
give $D_\mathrm{spot}\simeq 200\, \mathrm{km}^2\,\mathrm{s}^{-1}$ for $T=48~$h, 
consistent with the value obtained by the first method in Sect.~\ref{stattracers}.

\section{Conclusions and discussion\label{discussion}}

\subsection{Main conclusions}
In this paper, we have studied large-scale transport by convection flows in the solar photosphere 
from an observational perspective, using a Lagrangian passive tracer modelling approach to characterise
turbulent diffusion and the formation of dynamical structures and patterns affecting this transport process.

In Sect.~\ref{LCSanalysis}, we applied Finite Time Lyapunov Exponents (FTLEs) and Lagrangian Coherence Structure (LCS) 
analysis techniques to large field of views covering a significant fraction of the solar disc, for unprecedented
continuous periods of tracer tracking of up to six days. The large-scale maps we obtained reveal a clear and 
robust emergent dynamical LCS pattern tiling the solar surface isotropically at supergranulation scales. These
results provide a new global perspective, rooted in observational data, of the dynamical interplay between
order and chaos at the solar surface. In particular, by transiently accumulating particles and magnetic fields,
Lagrangian coherent structures appear to regulate large-scale turbulent surface diffusion in the quiet Sun
on long timescales. We also found that characterising their associated basic statistical properties (space 
and time scales) is sufficient to provide a correct preliminary dimensional order of magnitude estimate 
of the large-scale turbulent diffusivity.

A more quantitative statistical analysis of large-scale diffusion was presented
in Sect.~\ref{transport} using two different methods: a direct statistical analysis of tracers trajectories, 
which notably allowed us to pinpoint the role of "large-scale" flow kicks in driving the transport on long
timescales, and an analog to an ink spot molecular diffusion laboratory experiment.  Both analyses 
point to an effective horizontal turbulent diffusivity coefficient $D=2-3\times10^8~\mathrm{m}^2~\mathrm{s}^{-1}$
on the longest timescales of six days probed. We argued that an analysis on even longer times (much larger 
than the correlation timescale of supergranulation) would be desirable to make the convergence to an 
asymptotic, long-time statistical diffusion regime more apparent, but that such an analysis can not be 
easily conducted currently considering observational constraints.

\subsection{Comparison with previous work}
The passive tracers advection analysis presented in this paper bears some 
similarities with a previous analysis of this kind conducted by \cite{roudier09} 
on Hinode data on a 48~h timescale.  The authors derived a diffusion coefficient 
$D\simeq 430\,\mathrm{km}^2\,\mathrm{s}^{-1}$, broadly consistent with, but on the higher end 
of our own estimates.  It is possible that the timescale they probed was not long 
enough to obtain a converged long-time diffusion coefficient, with the statistics 
being dominated by shorter-time, superdiffusive ballistic advection within supergranules. 
The present analysis, conducted on times larger than the typical supergranulation lifetime 
$\tau_\mathrm{SG}$ and on significantly larger fields of views, offers a significant
statistical improvement in this respect.

Our results can also be compared with solar studies involving the tracking of 
magnetic elements. We argued that magnetic transport is overall well captured by the dynamics of 
virtual passive tracers/corks. The main benefits of the latter, we recall, is 
that they can be integrated for longer times, provide better statistics and do not
suffer from the issue of cancellation/merging effects \cite[see e.g.][for a discussion]{iida16}. 
As mentioned in the introduction, all studies
of this kind so far have focused on smaller fields of views and smaller timescales 
than the present study. Where the timescales of these studies overlap with ours, 
we find good agreement with our results. For instance,  \cite{hagenaar99} 
derived $D= 200-250\,\mathrm{km}^2\,\mathrm{s}^{-1}$ for magnetic tracking times 
in the $19-45$~h range; \cite{giannattasio13,giannattasio14} report a slightly hyperdiffusive 
regime with $D= 100-400\,\mathrm{km}^2\,\mathrm{s}^{-1}$ on supergranulation scales for a few hours.
Most other observational solar magnetic turbulent diffusion studies 
in the literature \citep[e.g.][]{berger98,utz10,abramenko11,manso11,jafarzadeh14,yang15,agrawal18} 
have been focused on intra-supergranular /internetwork transport on significantly 
smaller times ($10^3-10^5$~s) and spatial scales of a few megameters  at most
(see \cite{schrijver96,bellot19} for reviews). They typically obtain diffusion
coefficients smaller than $100\,\mathrm{km}^2\,\mathrm{s}^{-1}$ that likely correspond
to transport by shorter-lived, smaller-scale structures like granules or explosive granules,
with only a minor effect of weaker, but much longer-lived supergranulation-scale flows on such short
timescales.  These studies therefore differ in both scales, scope, goal and spirit from our study, 
which preferentially targets effective large-scale transport on times comparable to, 
or larger than the typical supergranulation lifetime $\tau_\mathrm{SG}=48~$h, 
which we believe is the appropriate limit to gauge the turbulent diffusion relevant 
to the large-scale solar dynamo problem.

\subsection{Implications for solar dynamo modelling}
Our long-time observational estimates of turbulent diffusivity, $D\simeq 2-3\times 10^8\,\mathrm{m}^2\,\mathrm{s}^{-1}$
obtained in Sect.~\ref{transport}, may be used to pin, at the solar photosphere, the profiles of horizontal 
turbulent magnetic diffusivity used for instance in mean-field models of the solar dynamo \citep{hazra23}.
If our results hold, some existing models might slightly underestimate this mode of 
transport \citep[e.g.][ who used $10^8\,\mathrm{m}^2\,\mathrm{s}^{-1}$ at the surface]{rempel06},
while others use a turbulent diffusivity value entirely consistent with our 
observational estimate (e.g. $D\simeq 2.5\times 10^8\, \mathrm{m}^2\,\mathrm{s}^{-1}$ in \cite{cameron10}).
On the other hand, some older flux-transport kinematic dynamo models \
\citep[][see \cite{schrijver96} for a discussion]{sheeley92,sheeley05}
seemed to require a significantly larger turbulent diffusion coefficient, up to
$D\simeq 6\times 10^8\, \mathrm{m}^2\,\mathrm{s}^{-1}$, to produce sensible solar dynamo
results. Such a large value is not consistent with even our upper-limit, 
enhanced transport estimates, $D\simeq 4\times 10^8\, \mathrm{m}^2\,\mathrm{s}^{-1}$,
obtained on the longest times probed, and associated with transiently re-accelerated "tail 
particles" (see inset of \fig{diffusionfig} and related discussion 
in Sect.~\ref{transport}). To be sustained on asymptotically long-times, 
such a large coefficient would require a significant 
transport regime change on longer timescales than the six days probed here. 
Such a dynamical effect is hard to envision at the surface at least, considering 
that our study already encompasses the main effects of supergranulation-scale 
convection, the most energetic large-scale flow structure contributing to turbulent
transport at the surface.

Further mean-field solar dynamo modelling work by \cite{lemerle15} (see their Fig.~8), 
highlighted in a recent review on surface flux transport by \cite{yeates23}, found a 
degeneracy between turbulent diffusivity and meridional flow amplitude parameters 
on the results of flux-transport dynamo models, which may somehow help to solve this
problem. We believe that the multi-pronged statistical characterisation of the 
horizontal turbulent diffusivity at the photosphere developed in the present paper 
is sufficiently robust to help lift this degeneracy: our 
result, $D\simeq 2-3\times 10^8\, \mathrm{m}^2\,\mathrm{s}^{-1}$, is very close 
to the turbulent diffusivity value associated with the maximum fitness contours 
obtained by \cite{lemerle15} in the turbulent diffusivity-meridional flow amplitude 
plane.

Finally, we note that transport coefficients derived from global numerical simulations 
of solar MHD convection also seem consistent with our observational estimates
(e.g. Fig.~11 of \cite{simard16} for the $\beta_{\varphi\varphi}$ transport 
coefficient corresponding to our observational measurement of the horizontal diffusivity),
although they might still slightly underestimate the vigour of the convection at 
supergranulation scales, and the associated diffusivity.

\subsection{Connection with the solar magnetic network, atmosphere and solar wind}
It is tempting to conjecture that the prominent global emergent dynamical pattern 
tiling the surface of the Sun, singled out in this work and best illustrated  
by \fig{FTLEglobalfig}, plays a major role in structuring the magnetic interactions 
between the interior, or at least the subsurface layers of the Sun driving 
the small-scale surface dynamo, its atmosphere, and possibly also the solar wind. 
The possible connection of the LCS loci of photospheric 
magnetic field accumulation with coronal heating has already been pointed out by \cite{yeates12}. 
Our new global characterisation reveals the conspicuity, robustness and full extent 
of this dynamical LCS pattern over the whole solar disc, reinforcing this hypothesis. 
In this respect, our analysis may also provide new insights into the origin and mechanisms 
underlying a possible connection, suggested by \cite{fargette21}, between 
supergranulation and solar wind magnetic switchbacks encountered by the Parker Solar 
Probe \citep{kasper19,bale19}.

\subsection{A new "experimental" measurement of turbulent transport}
Moving on to a more fundamental complex physics and fluid-dynamical perspective, we note that
turbulent diffusion effects have been observed
in various experimental MHD flows, such as the Perm, Wisconsin and VKS dynamo experiments 
\citep{frick10,forest12,miralles13}, all at relatively large kinetic Reynolds numbers $Re=O(10^6)$
but rather small magnetic Reynolds numbers ($Rm<100$) not representative of astrophysical regimes
such as encountered in planets, stars and galaxies.
By using "the Sun as a fluid dynamics lab", our results offer a new, and possibly 
unique "experimental", \textit{in situ} caracterisation of the emergence on times comparable to, 
or longer than a typical flow correlation time, of turbulent diffusion in stochastic flows 
with both large kinetic and magnetic Reynolds numbers ($Re=O(10^{10})$, $Rm=O(10^4)$ at the solar surface).

\subsection{The pitfalls of mixing-length arguments}
Looking at the results from a broader perspective than just the solar context to benefit
from the insights of the study of an astrophysical system resolved in both space and time, 
it is also an interesting exercise to ask why the value of the turbulent diffusivity we obtained is 
what it is in this system. If the transport is indeed dominated by supergranulation-scale 
flows, na\"ively (by a mixing length argument) one could have expected a transport coefficient 
of the order of a fraction (typically $1/4$ in 2D) of $R_\mathrm{SG} V_\mathrm{SG}$, i.e. 
\begin{equation}
\label{mixinglengtheq}
D=\f{1}{4} R_\mathrm{SG} V_\mathrm{SG}= \f{1}{4}\f{R_\mathrm{SG}^2}{\tau_\mathrm{NL,SG}} = 1.5\times 10^9\,\mathrm{m}^2\,\mathrm{s}^{-1}~,
\end{equation}
where we have used $R_\mathrm{SG}\simeq 1.5\times 10^4\,\mathrm{km}$,
and $V_\mathrm{SG}\simeq 0.4\,\mathrm{km}\,\mathrm{s}^{-1}$ \citep{rincon18}, and we have 
introduced the nonlinear turnover time
at the supergranulation scale $\tau_\mathrm{NL,SG}=R_\mathrm{SG}/V_\mathrm{SG}\simeq 8~$h, which
is the sweeping time for a passive tracer to be advected from the centre to the boundary of a supergranule cell.
This mixing-length estimate is an order of magnitude higher than our experimental-observational determination.
Why is it so ? The reason lies in that we have na\"\i vely used, in the expression above, 
the turnover time of the flow $\tau_\mathrm{NL,SG}$, instead of its correlation time, $\tau_\mathrm{SG}$, 
which is closer to 48~h. If we repeat the calculation in \equ{mixinglengtheq} with this time instead, we find
$D=2.6\times 10^8\,\mathrm{m}^2\,\mathrm{s}^{-1}$, which is now remarkably consistent with our detailed statistical
analysis.

The reason why the correlation time of the flow is the relevant quantity to use in a mixing length estimate 
here is because tracers remain stuck at the boundary of supergranules for a time of the order of $\tau_\mathrm{SG}$
after they have been advected there on the much shorter $\tau_\mathrm{NL,SG}$ time. 
Hence, their effective "random walk" velocity is not $V_\mathrm{SG}=0.4\,\mathrm{km}\,\mathrm{s}^{-1}$, 
but the much smaller $V_\mathrm{SG}\times (\tau_\mathrm{NL,SG}/\tau_\mathrm{SG})\simeq 70\,\mathrm{m}\,\mathrm{s}^{-1}$.
This simple, yet subtle difference finds its roots in the very structure of the flow.

The conclusion of this phenomenological argument is therefore that the structure of a turbulent flow
matters a lot when it comes to correctly estimating the magnitude of large-scale turbulent transport. 
In the example at hand, we showed that the global surface network of transport barriers and LCS at 
supergranulation scales, vividly illustrated in \fig{FTLEglobalfig}, plays a key role 
in the regulation of the effective large-scale transport. We believe that this conclusion 
pertains to many if not all astrophysical flows, and therefore call for caution with
back-of-the-envelope estimates of turbulent transport coefficients based on dimensional arguments, 
for instance because the correlation time of the flow can significantly differ from 
its turnover time. In the case of solar surface convection, large-scale structures 
at the injection scale persist for quite a long time, so that the Strouhal number of the 
flow, $\mathrm{St}=\tau_\mathrm{corr}/\tau_\mathrm{NL}$, is 
of the order of 5-6, with significant consequences for the effective turbulent diffusivity.

\subsection{Perspectives}
The focus of this paper has deliberately been restricted to a subset of all
physical solar dynamical transport and large-scale organisation
phenomena to which the techniques developed in this work may be applied 
at global scales. Further investigations of this kind, of possible rotational effects, latitudinal 
dependences of turbulent diffusivity, global-scale convection \citep{hathaway13,ballot24}, 
meridional circulation \citep{roudier18}, and of the statistical implications of magnetic 
flux emergence and further transport at the photosphere \citep{hathaway12} 
for the energetics of the upper solar layers \citep{yeates12}, would undoubtedly 
prove very instructive too, and are left for future work. 

More broadly, as exemplified by the discussion above, the study of the extreme 
fluid dynamical system that is the Sun, with the unique benefits in astrophysics of a 
large spatial and temporal resolution, can still teach us valuable lessons for the future modelling 
of unresolved astrophysical systems, such as stars or accretion discs, where turbulent flows 
driven by different hydrodynamic or MHD instabilities also likely play a key role 
in the overall dynamical and energetic regulation of the system.

\begin{acknowledgements}
We thank Peter Haynes and Yves Morel for several interesting
conversations and for their invitation to discuss
and present a preliminary version of the results at the 2019 TEASAO
workshop in the beautiful surroundings of the Saint-F\'err\'eol
lake, and Michael Nastac for a useful discussion
on the theory of anomalous transport scalings in turbulence during an astrophysical
plasma workshop organized at the Wolfgang Pauli Institute in Vienna, whose hospitality 
is gratefully acknowledged. We also thank Nathana\"el Sch{\ae}ffer for his assistance 
with the \texttt{SHTns} library \citep{schaeffer13}, the SDO/HMI data provider JSOC and
the HMI/SDO team members for their hard work. In particular, we are
grateful to P. Scherrer, S. Couvidat and J. Schou for sharing
information regarding the calibration and removal of systematics of
HMI Doppler data.  This work was granted access to the HPC resources of 
CALMIP under the allocation 2011-[P1115].  We thank Nicolas Renon for his 
assistance with the parallelization of the CST code. This work
was supported by COFFIES, NASA Grant 80NSSC20K0602.
\end{acknowledgements}
\bibliographystyle{aa}
\bibliography{sg}

\appendix

\section{Computation of Finite-Time Lyapunov exponents\label{FTLEapp}}
\subsection{Computation of Lagrangian tracers trajectories\label{tracersFTLEapp}}
The trajectories of Lagrangian tracers (``corks'') in the $(x,y)$
plane are determined as follows: a cartesian grid of tracers is
initially generated in the plane. The time history of the $(x,y)$
projection of the photospheric velocity field
$\vec{u}(x_i,y_j)=(u_x(x_i,y_j),u_y(x_i,y_j))$, determined via the CST
algorithm every 30~min, is then used to integrate their position in
that plane for a target time $T$ ranging
from several hours to several days, by means of numerical integration of an
advection equation implemented through the python function \texttt{odeint}. 
The integrator internally sets an adaptative timestep
shorter than the time between successive velocity snapshots. At each such
timestep, the velocity field at the exact current location of each
tracer is interpolated using the running  snapshot of the velocity
field on the cartesian grid, thanks to the
python function $\texttt{RectBivariateSpline}$. 
Note that using the same velocity field snapshot during
30~min intervals is a reasonable assumption, given that
the smallest spatial scales of the photospheric flow inferred via the
CST are of the order of 2.5~Mm, and their timescale is of the order of
30~min to 1~hour. Note also that the resolution of the tracers grid on
which a FTLE field is to be computed can be
finer than the flow itself. Using a high-resolution grid of
tracers is in fact essential to identify Lagrangian coherent
structures and accurately compute invariant manifolds and transport
barriers even when the flow is large-scale \citep{lekien05}. 
For global maps, we used a cartesian grid of up to $1024\times 1024$
tracers encapsulating the full solar disc. This maximal resolution 
corresponds to an initial spacing between tracers of
the order of 1.5~Mm close to the disc centre. The precision of the
integration of the tracers trajectories, specified to the
numerical integrator, is of the order of 0.3~Mm (in Sect.~\ref{transport} 
we also used a much finer local grid of $1024\times 1024$ tracers encompassing
$10\degree\times 10\degree$ only). The collection of trajectories integrated in the CCD plane 
then serves as a basis for the computation of FTLE on the sphere, as explained below.

\subsection{FTLE in a 2D plane\label{2DFTLEapp}}
The Lagrangian trajectory $\vec{x}(t)=(x(t),y(t))$ of a passive tracer
in a 2D plane fluid flow $\vec{u}(\vec{x},t)$ is determined
by its initial position $\vec{x}(t=0)=\vec{x}_0$ and the differential
equation of motion
\begin{equation}
\label{eq:trajectory}
\dot{\vec{x}}(t)=\vec{u}(\vec{x}(t),t).
\end{equation}
Introducing the flow map for a target integration time $T$
\begin{equation}
\begin{array}{cl}
\phi^T: & \mathbb{R}^2\rightarrow \mathbb{R}^2\\
& \displaystyle{\vec{x}_0 \mapsto \vec{x}(T,\vec{x}_0)}~,
\end{array}
\end{equation}
and its Jacobian $d\phi^T/d\vec{x}_0$, we form the Cauchy-Green
deformation tensor
\begin{equation}
\label{eq:cauchy}
\Delta(\vec{x}_0)=\left(\f{d\phi^T}{d\vec{x}_0}\right)^\tens{T}\left(\f{d\phi^T}{d\vec{x}_0}\right)~.
\end{equation}
The FTLE $\sigma^T(\vec{x}_0)$ of the flow at $\vec{x}_0$ is given by
\begin{equation}
\label{eq:lyapunov}
\sigma^T(\vec{x}_0)=\f{\ln\sqrt{\lambda_{\mathrm{max}}(\Delta(\vec{x}_0))}}{T},
\end{equation}
where $\lambda_{\mathrm{max}}$ is the largest eigenvalue of
$\Delta(\vec{x}_0)$.

In practice, the matrix representation $\tens{J}$ of $d\phi^T/d\vec{x}_0$ is
estimated at each interior point $(x_i,y_j)$ of the cartesian grid on
which the tracers are placed at $t=0$, using a centered finite
difference formula.  Using the short-hand notation
  $\vec{x}_{i,j}(t)\equiv\vec{x}(t,\vec{x}_0=(x_i,y_j))$, we have
\begin{equation}
\label{eq:differentiationmatrix}
\tens{J}(x_i,y_j)=\left(
\begin{array}{cc}
\displaystyle{\frac{x_{i+1,j}(T)-x_{i-1,j}(T)}{x_{i+1,j}(0)-x_{i-1,j}(0)}} &
\displaystyle{\frac{x_{i,j+1}(T)-x_{i,j-1}(T)}{x_{i,j+1}(0)-x_{i,j-1}(0)}}
\\ \\
\displaystyle{\frac{y_{i+1,j}(T)-y_{i-1,j}(T)}{x_{i+1,j}(0)-x_{i-1,j}(0)}} & \displaystyle{\frac{y_{i,j+1}(T)-y_{i,j-1}(T)}{x_{i,j+1}(0)-x_{i,j-1}(0)}}
\end{array}
\right)~,
\end{equation}
from which $\Delta=\tens{J}^\tens{T}\tens{J}$ follows. The FTLE field
$\sigma^T(x_i,y_j)$ is finally obtained via \equ{eq:lyapunov} by
diagonalizing $\Delta$ at each grid point.

\subsection{Mapping to the sphere\label{mappingFTLEapp}}
The FTLE of the flow on the sphere is inferred using basically
the same algorithm presented above, except that a
mapping between the real separations between the tracers and
the projected ones in the plane of the sky/satellite CCD must be
introduced. The theoretical formalism is
described in \cite{lekien10}, as well as an example of
application on the sphere. The formulae needed to compute
FTLEs on the sphere for our particular problem are given below. 

To perform the mapping, we introduce the out-of-plane
distance $z$ between a point on the solar surface and the plane
parallel to the CCD plane and passing through the centre of the Sun, 
\begin{equation}
z=\sqrt{\rsun^2-(x^2+y^2)}~,
\end{equation}
where $\rsun$ is a fiducial photospheric solar radius. The
transformation between the solar disc $D_\odot$ and spherical
solar surface manifold $\mathcal{M}$ is then 
given by the diffeomorphism
\begin{equation}
\label{eq:chart}
\begin{array}{cl}
\beta^{-1}: & D_\odot \rightarrow \mathcal{M} \\
           & (x,y) \mapsto \left(\begin{array}{c} x\\ y \\
               z=\sqrt{\rsun^2-(x^2+y^2)}\end{array}\right)~.
\end{array}
\end{equation}
Using these definitions, the FTLE field on the sphere is obtained
along the exact same lines as in Appendix~\ref{2DFTLEapp}, except that
a more general form of the deformation tensor accounting for the 
projection effects must be used, namely
\begin{equation}
  \label{eq:cauchysphere}
\Delta(\vec{x}_0)=\widetilde{\tens{J}}(\vec{x}_0)^\tens{T}\,\,\widetilde{\tens{J}}(\vec{x}_0)~,
\end{equation}
with
\begin{equation}
\label{eq:Jtilde}
\widetilde{\tens{J}}(\vec{x}_0)=\tens{R}\left(\vec{x}(T,\vec{x}_0)\right)\,\tens{J}(\vec{x}_0)\,\tens{R}\left(\vec{x}_0\right)^{-1}~.
\end{equation}
$\tens{J}$ is the Jacobian matrix in the projection plane,
computed at each grid point via \equ{eq:differentiationmatrix}
using the projected tracers trajectories. $\tens{R}(\vec{x})$ is a coordinate-dependent
upper-triangular matrix obtained by QR decomposition of the derivative
of $\beta^{-1}$. Its explicit expression for our particular problem is
\begin{equation}
\tens{R}(\vec{x})=%
\left(%
\begin{array}{cc}
\displaystyle{\sqrt{1+ \f{x^2}{z^2}}} & \displaystyle{\f{xy}{\left(z\sqrt{\rsun^2-y^2}\right)}} \\
 & \\
0 & \displaystyle{\f{\rsun}{\sqrt{\rsun^2-y^2}}}
\end{array}
\right)~.
\end{equation}
Note that since the particles move on the sphere between 
$t=0$ and the target time $t=T$, the projection effects at $T$ are
different from those at the initial time. This explains why $\tens{R}$
is evaluated at the final position $\vec{x}(T)$ on the left of
$\tens{J}$ in \equ{eq:Jtilde} and at the initial position $\vec{x}(0)$ on its
right.

\end{document}